\documentclass[12pt,leqno]{article}
\usepackage{amssymb,amsthm,amsmath}
\usepackage{graphicx}
\usepackage{fullpage}
\usepackage{color}
\usepackage{epstopdf}
\usepackage[blocks]{authblk}

\newtheorem{defin}{Definition}
\newtheorem{definition}{Definition}
\newtheorem{proposition}{Proposition}

\def\PE{{\rm PE}} 
\def\TH{\ell} 
\def\MIR{{\rm MIR}} 
\def\POL{{\frak P}}

\def\MaxJ{\frak{J}}
\def\Hyp{H}
\def\Prob{\mathbb{P}}
\def\Score{{\rm Score}}
\def\Esp{\mathbb{E}}
\def\TO{\mathbf{T}}
 
\def\mp{{\frak m}}

\def\one{\mathbf{1}}

\def\R{\mathbb{R}}
\def\X{\mathbb{X}}

\def\Card{\#}
\title{Realtime market microstructure analysis: online Transaction Cost Analysis}
\author[1,2]{\small R. Azencott}
\affil[1]{\small Department of Mathematics, University of Houston, Houston, TX.}
\affil[2]{\small Emeritus Professor, Ecole Normale Sup. Cachan, France.}
\author[3]{A. Beri}
\affil[3]{\small Mathematical 
Biosciences Institute, The Ohio State University, Columbus, OH.} 
\author[1]{Y. Gadhyan}
 \author[4]{N. Joseph}
\affil[4]{\small Cheuvreux, Cr\'{e}dit Agricole Group, Paris, France}
\author[4]{C. A. Lehalle}
\author[4]{M. Rowley}
\begin{document}
\maketitle

\begin{abstract}
Motivated by the practical challenge in monitoring the performance of a large number of algorithmic trading orders, this paper provides a methodology that leads to automatic discovery of causes that lie behind poor trading performance. It also gives theoretical foundations to a generic framework for real-time trading analysis. 
The common acronym for investigating the causes of bad and good performance of trading is TCA (Transaction Cost Analysis \cite{citeulike:10211148}).
Automated algorithms take care of most of the traded flows on electronic markets (more than 70\% in the US, 45\% in Europe and 35\% in Japan in 2012). Academic literature provides different ways to formalize these algorithms and show how optimal they can be from a mean-variance \cite{OPTEXECAC00}, a stochastic control \cite{citeulike:9304794}, an impulse control \cite{citeulike:5797837} or a statistical learning \cite{citeulike:10160160} viewpoint. This paper is agnostic about the way the algorithm has been built and provides a theoretical formalism to identify in real-time the market conditions that \emph{influenced} its efficiency or inefficiency.
For a given set of characteristics describing the market context, selected by a practitioner, we first show how a set of additional derived explanatory factors, called \emph{anomaly detectors}, can be created for each market order (following for instance \cite{citeulike:522801}). We then will present an online methodology to quantify how this extended set of factors, at any given time, predicts (i.e. have \emph{influence}, in the sense of predictive power or information defined in \cite{cristianini2000introduction}, \cite{shannon2001mathematical} and \cite{alkoot1999experimental}) which of the orders are underperforming while calculating the predictive power of this explanatory factor set.
Armed with this information, which we call \emph{influence analysis}, we intend to empower the order monitoring user to take appropriate action on any affected orders by re-calibrating the trading algorithms working the order through new parameters, pausing their execution or taking over more direct trading control. Also we intend that use of this method in the post trade analysis of algorithms can be taken advantage of to automatically adjust their trading action.
\end{abstract}
\paragraph{Acknowledgements.}
\textit{This research has been supported by the Cr\'edit Agricole Cheuvreux \emph{Trading and Microstructure} Research Initiative, in partnership with the Louis Bachelier Institute and the Coll\`ege de France.}

\tableofcontents

\section{Introduction}

Institutional investors use optimal dynamic execution strategies to trade large quantities of stock over the course of the day. Most of these strategies have been modelled quantitatively to guarantee their optimality from a given viewpoint: it can be from an expectation \cite{BLA98}, a mean-variance \cite{OPTEXECAC00}, a synchronized portfolio \cite{citeulike:5094012} or a stochastic control \cite{citeulike:5797837} perspective.

Algorithmic trading analyzes high frequency market data, viewed as actionable information, in order to automatically generate and place automated trading orders. Hence, automatic trading algorithms can be viewed as parametrized black boxes. 
They are initialized and monitored by human traders, who have the capability to set and adjust some high-level parameters that drive the algorithm - incorporating the a trader's view and allowing reaction to unexpected events. 
However standard automatic trading systems, do not currently offer advanced computerized discovery of potential causes for poor trading performance.
Our paper focuses on automated online monitoring of portfolio performance by real-time scanning of static (like the sector, the country, etc, of the traded stock) and dynamic (like the bid-ask spread, the volatility, the momentum, the fragmentation of the traded stock) ``explanatory'' market variables quantifying their current influence on portfolio performance. The \emph{Influence Analysis} methodology we have developed and tested can provide real-time feedback to traders by detecting the most significant explanatory factors influencing current degradations of trading performance on specific portfolios.

TCA practitioners, in particular for real-time analysis, are faced with the problem of automating the understanding of how market context affects trading performance of a large number of orders.
This paper is the first theoretical formalization of such a process proposing a framework to understand and improve TCA (off line or in realtime) in several aspects: (1) augmenting the description of the market context (using scoring --section \ref{sec:scores}-- and pattern detection --section \ref{detectors}--) to identify relationships between this enhanced description and the performance of a large basket of orders.
A key feature of our influence analysis methodology is that it does not require actual knowledge of the trading algorithms mechanisms. Online influence analysis could hence be useful in updating trade executions as well as in re-calibrating trading algorithms. Influence analysis on historical data could also improve existing trade scheduling algorithms, selecting new significant signals in their kinematics.

Our online influence analysis generates in real-time a very short list of market factors ``explaining'' the current lack of performance in a collection of intra-day trades. The automatic trading algorithms can be arbitrary; 
it can be a result of a classical mean-variance optimization (see \cite{OPTEXECAC00}), stochastic control of a brokerage benchmark (see \cite{citeulike:5797837}), stochastic control of a market making scheme 
(see \cite{citeulike:9304794}), a liquidity-seeking algorithm driven by a stochastic algorithm optimization (see \cite{citeulike:10160160}), or even purely heuristic driven ones. 
The analysis conducted here is supported by automatic detection of conjunction of poor trading quality and singularity or rarity of the market context. The influence analysis can be launched as soon as a portfolio performance evaluation criterion is selected, such as the slippage with respect to the arrival price, to the VWAP (volume-weighted average price), etc.
Our online influence analysis relies on extensions of  classical relative entropy techniques (see for instance \cite{brillinger2004some} \cite{mougeot2011traffic} \cite{azencott4analysis} \cite{billingsley1965ergodic}) to generate in real-time optimized empirical relations between an automatically selected small set of high influence explanatory factors and any pre-assigned trading performance criterion. Our approach also quantifies at each time $t$ the current influence of a given pool of market factors on trading performance degradation. 

We first describe typical sets of market variables and trading performance criteria to which our influence analysis applies, and we outline our benchmark sets of intra-day data, used to test our approach.
In section \ref{input} we describe the market dataset considered for our study. 
We then present in section \ref{detectors} the three online anomaly detectors we have developed to enrich in real-time any set of dynamic input market variables.
Such detectors are bespoke implementation of a more generic class of detectors that could be used like wavelet coefficients \cite{citeulike:3974506}.
Section \ref{influence} outlines our generic framework for influence analysis. We select and fix any pragmatic binary trading performance criterion $Y_t$ detecting low trading performance at time $t$. This binary criterion will be deduced from a variable of performance $\PE_t$ of the trading portfolio. Given any small set $G$ of explanatory variables $\X^G = \{X_t^j, j \in G\}$ (deduced from a multiscale analysis of market context descriptors $\{M_t^j,1\leq j\leq J\}$), we generate the current "best" predictor $\hat Y_t = h_t( \{X^G\} )$ of $Y_t$ based on these explanatory variables, and we compute its current predictive power, which we call    the influence coefficient $J_t(G)$ of the group $G$ on $Y_t$   
The time dependent set $G_t^{\max}$ which maximizes  $J_t(G) $ among all small groups $G$ of explanatory factors  is then determined, and if $J_t(G_t^{\max})$ is high enough, the set $G_t^{max}$ can be exported in real-time to traders, as the set of market variables which best explains current trading  performance degradations. 
In section \ref{compute.influence} we present the accuracy analysis of the influence computation, and obtain pragmatic conditions for robust identification of explanatory variables. We then present the steps to compute optimized parametric predictors based on single explanatory factors, and to generate  hierarchical combinations of optimal predictors. 
Section \ref{output} presents the test results of our influence analysis on benchmark intra-day datasets provided by Cr\' edit Agricole Cheuvreux Quantitative Research Group.

\section{ Dynamic dataset to be monitored online} \label{input}
\subsection{The automated trading process}\label{sec:tprocess}

Only few assumptions are demanded to a trading process to be monitored by the methodology proposed here. By ``\emph{trading process}'' we mean the operation of buying or selling shares or any other financial instrument in more than one transaction.
Since the ``last leaf'' of any investment or hedging strategy is to obtain transactions, a trading process is needed if a block execution cannot be obtained.
With the conjunction of the financial crisis and regulation changes (mainly Reg NMS in the US in 2005 and MiFID in Europe in 2007, see \cite{citeulike:12047995} for more details), the capability to close a deal in one transaction strongly decreased. Hence most market participants are ``\emph{slicing}'' their large orders (see \cite{OPTEXECAC00}, \cite{citeulike:5797837} or \cite{citeulike:5177512} for quantitative approaches of optimal slicing).
At a $\delta t$ minutes time scale (say $\delta t=5$), an important variable of a trading process is its ``\emph{participation rate}'' $\rho_{m\cdot\delta t}$; the trader (manually or tuning parameters of some trading robots), succeeds in obtaining $\rho_{m\cdot\delta t} \, V_{m\cdot\delta t} $ shares during the $m$th interval of $\delta t$ minutes when the whole market traded $V_{m\cdot\delta t} $ shares.

The adequate trading rate is difficult to achieve. First because its optimal value is a function of market conditions during the whole trading process (i.e. it is not causal). Secondly, market conditions are either price driven (like the volatility, the presence of price jumps or trends, correlations between some assets, etc) or liquidity driven (the bid-ask spread, the market depth, etc); that all of these are difficult to anticipate or predict.
Moreover, even if the theoretically optimal trading rate would have been known in advance, the actions of the trader himself have an impact on future market conditions \cite{citeulike:9771410}. A trader is thus continuously monitoring his trading rate in conjunction with market conditions to check: first that ex-post his trading rate has been close to the expected one, second that a change in his trading rate does not come from an unexpected change of market conditions, and last that he is not impacting the price formation process.
The faster he understands what is happening to the trading process, the more efficiently he will be able to react.

The motivation to buy or sell does not change the way to apply the ``\emph{influence analysis}'' methodology presented in this paper. It can be driven by long term considerations (typical for orders sent by institutional investors to executing brokers' trading desks), to hedge a derivative portfolio, to implement an arbitrage strategy, or even inside a market making scheme (see \cite{citeulike:9304794}).
For all of these motivations the same trading process takes place: a human being monitors a trading rate according to performance criteria (that are specific) and tries to adjust his trading rate as fast as possible to take into account changes in market conditions.
The methodology proposed in this paper provides decision support to the trader. 

\paragraph{Performance criteria of a trading process.}
In our framework, the proxy for the target of the trader is called his \emph{performance evaluation criterion}. Let us define a few possible criteria:
\begin{itemize}
\item for a market-maker: a decreasing function of his inventory imbalance, and his profits are good performance criteria;
\item for a brokerage trading desk: a decreasing function of the distance to a fixed participation rate (for instance 10\% or 20\% of the market traded volume), the obtained average price compared to the Volume Weighted Average Price (VWAP) of the market, or to the arrival price, or to the close price, are meaningful criteria;
\item for an arbitrageur: a decreasing function of his tracking error, and his profits should be chosen.
\end{itemize}

\paragraph{Description of the market context.}
In addition to the performance evaluation criteria, we will need \emph{market descriptors} to quantify the market context. They will be used by the proposed methodology to build an online understanding of the causes of bad performance. Typical market descriptors are:
\begin{itemize}
\item \emph{Price driven market descriptors}:
  \begin{itemize}
  \item prices returns or price momentum (signed to be in the same direction as the side of the monitored order) are important since it is more difficult to buy when the price is going up rather than when it is going down.
  \item The volatility is a common proxy for the amount of uncertainty in the price dynamics. Its influence on trading performances is not straightforward since some volatility can help to capture passively some flows, but a too high volatility level can lead to adverse selection. Moreover, market impact models link the volatility to the impact of a trade a negative way (see \cite{citeulike:4325901}), meaning that a too high level of volatility is negative for almost all trading processes.
  \end{itemize}
\item \emph{Liquidity driven market descriptors}:
  \begin{itemize}
  \item the bid-ask spread is the distance (in basis points of the current price) between the best ask price and the best bid one. It thus describes the state of the ``auction game'' between market participants (see \cite{citeulike:10363469} for more details and terminology).
  \item Market traded volume is an important characteristic of the market since it is easier to buy or sell when the market is active than when nothing is traded.
  \item The visible size at first limits (also called \emph{average volume on the books}) can also be used to quantify the current \emph{market depth}.
  \end{itemize}
\end{itemize}

Any other characteristic of the trading instruments can be added in the analysis. For shares, modal variables like the sector of the stock or the market capitalization of the listed firm are of importance.

\paragraph{Anomaly detectors.}
To be able to capture the causes of bad trading performance, it is often useful to have access to information other than the averaged values of market variables. For instance price jumps, price trend changes, volume peaks and crenels are not captured by averages and we would like to take them into account in our \emph{influence analysis} methodology. Section \ref{detectors} shows how to build such detectors that will be used in the analysis.

\subsection{Trading orders }

We consider a portfolio of at most $K$ trading orders $\TO(k), k= 1, \ldots, K$ driven by automatic trading algorithms, and supervised by one or more traders. Each trading order is defined by a few ``static'' variables such as buy/sell label, order size, trading place, section, country, capitalisation, free-float, benchmark type (VWAP, arrival price, etc), etc. 
In our intra-day benchmark studies a portfolio typically involves $200 \leq K_t \leq 700$ active orders at any arbitrary 5 minute time slice.
%
\subsection{Market descriptors}
\label{sec:scores}

Each trading order $\TO(k)$ focuses on a specific asset whose dynamics is recorded at each time point $t$ through a fixed number of basic ``market descriptors'' $M^1_t(k), M^2_t(k) , M^3_t(k), \ldots $; in our benchmark study below, we have focused on a subset of  the following  market descriptors:
\begin{itemize}
\item $M^1 =$ Volatility
\item $M^2 =$ Spread
\item $M^3 =$ Momentum in bid-ask spread
\item $M^4 =$ Momentum in bp
\end{itemize}
This list  can be augmented by the  \emph{rarity scores} $\Score( M^i)$ of the market variables $M^i$. These scores are defined by  $\Score( M^i) = F^i(M^i)$ where  $F^i$ is the cunulative distribution function of  $M_i$. Each such  score  necessarily has a uniform distribution \cite{Borovkov98}. If  poor performance on a given set of stocks is due to a strong increase in  the volatility level,   the concrete cause may either be due to volatility reaching an "absolute"  psychological threshold, or to volatility being high  \emph{relatively to its usual levels}. In this last case the volatility score will be a better explanatory factor for poor performance.
We will  use here the following scores, increasing our number of market variables:
\begin{itemize}
\item $M^5 =$ Volume Rarity Score
\item $M^6 =$ Volatility Rarity Score
\item $M^7 =$ Spread Rarity Score
\end{itemize}

Figure \ref{fig:wimap} {displays typical intraday} (January 14, 2011) 
plots of the time series corresponding to the 7 basic market variables listed above for an anonymous stock.

\begin{figure}[h!]
\centering
\includegraphics[width=\linewidth]{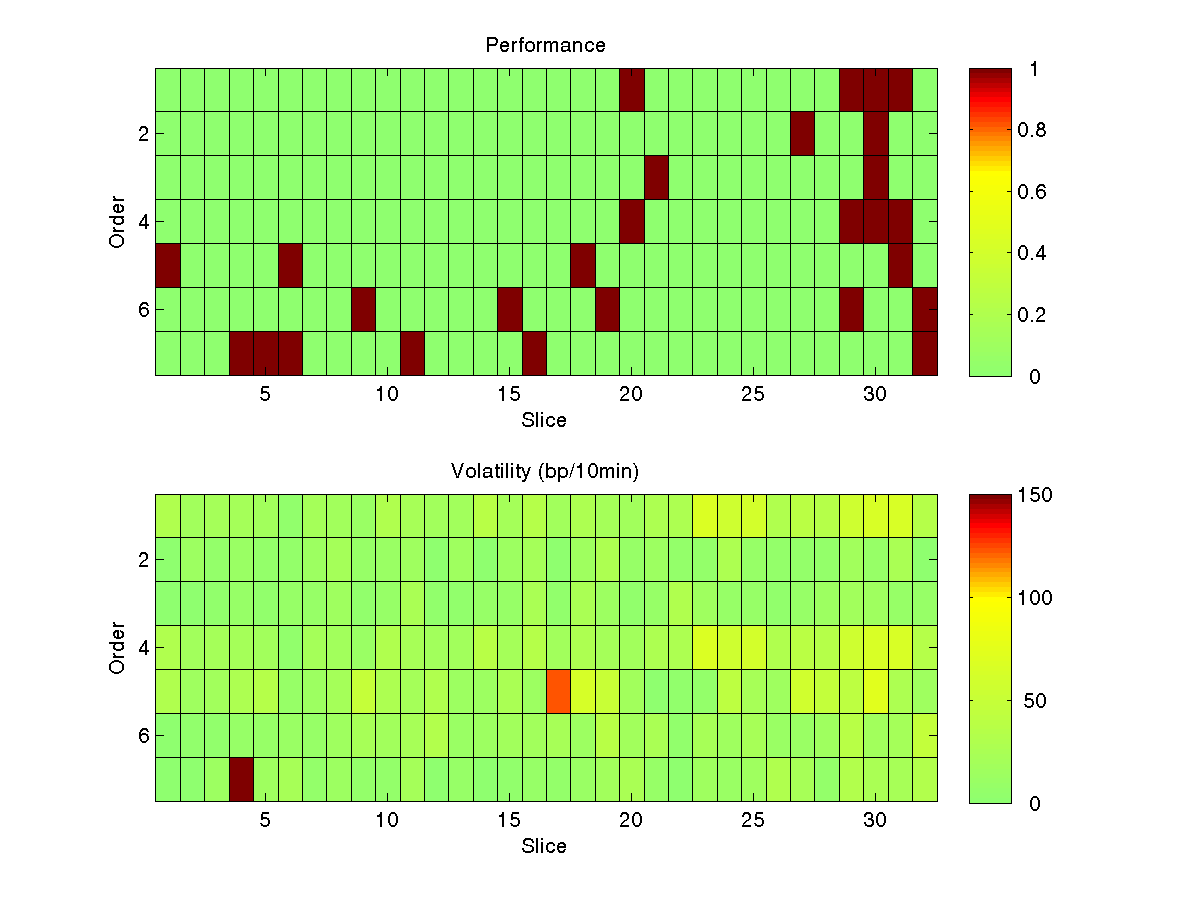}
\caption{Heatmap of online performances of a traded basket of 7 orders (top) in conjunction with values of one explanatory variable (the volatility; bottom), the correspondance between the two heatmaps is not obvious.The abcissa represents the evolution over five-minute time slices.}
\label{fig:perfheatmap}
\end{figure}

We also display heat map representations of rarity scores for multiple stocks in Figures \ref{fig:perfheatmap} and \ref{fig:perfheatmapscore}, where each row displays the time series of rarity scores for one single stock (associated to one trading order in our benchmark data), and each column represents one time slice. Clearly ``extreme'' rarity scores tend to appear in clusters, and to co-occur across multiple stocks.

As will be seen below, high co-occurrence frequency of ``dynamic anomalies'' such as peaks, jumps, etc within a group of stocks tend to ``explain'' simultaneous lack of performance for the corresponding trading orders. 
\begin{figure}[h!]
\centering
\includegraphics[width=\linewidth]{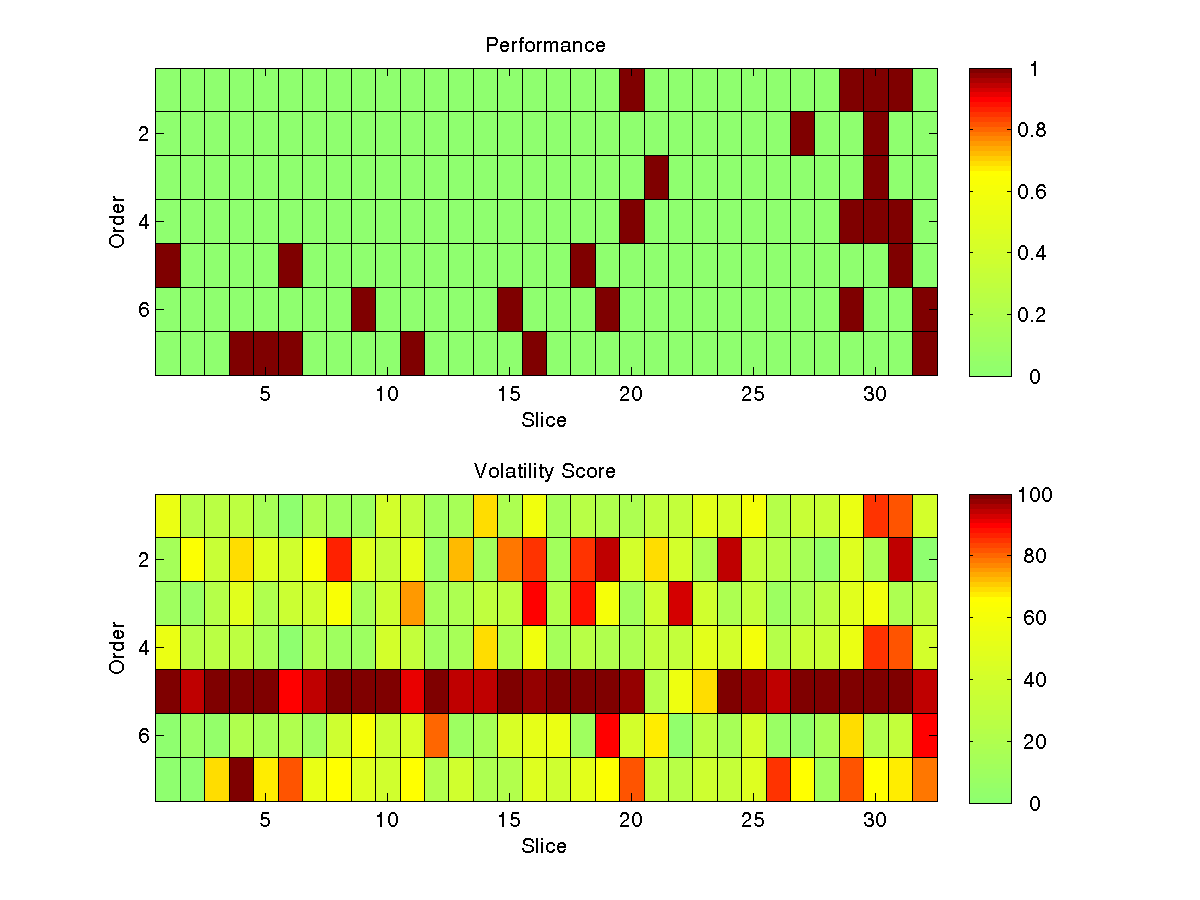}
\caption{Heatmap of online performances of the some basket of 7 orders (top) in conjunction with the scores of the same explanatory variable (the volatility; bottom), compared to Figure \ref{fig:perfheatmap}, one can see that the scores change the high and low values of the explanatory variables, giving birth to more potential conjunctions with bad trading performances.}
\label{fig:perfheatmapscore}
\end{figure}
\subsection{Trading performance evaluation}

We also select a ``trading performance evaluator'' $\PE{}$ providing at each time point $t$, and for each active trading order $\TO(k)$, a quantitative evaluation $\PE{}_t(k)$ for the current performance of $\TO(k)$. 

In our study we have selected by default $\PE{} = $ ``slippage in bid-ask spread'' (\emph{slippage} being the average price of the order minus the benchmark --VWAP, arrival price, close price, etc-- for a sell order, and the opposite for a buy order), but there are no restrictions on the user choice for this $\PE$ variable. In particular, other examples of $\PE{}$ include ``slippage in bp'' (basis points), ``Slippage in Dollars'', ``Absolute value of slippage in bid-ask spread'', etc. 
We are assuming that degraded performances are associated to low values of $\PE{}$.
For each trading order $\TO(k)$ the performance evaluator and the 7 market descriptors are volume averaged over successive time slices of arbitrary duration (set at 5 minutes for our benchmark study). Thus we generate 8 time series $M^1_t(k), \ldots , M^7_t(k)$ and $\PE{}_t(k)$ indexed by time slices $t$.
These time series generically have missing values since orders do not necessarily begin or end at the same time.
Fix a low percentile threshold  $q$ such as $q= 5 \%$ or $q= 3 \%$ to binarise the performance evaluator.

\begin{figure}[h!]
\centering
\includegraphics[width=.6\linewidth]{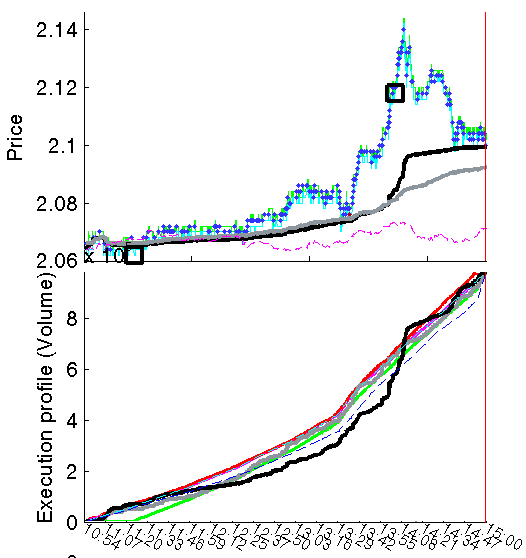}
\caption{Example of intraday behaviour of performance variables for a given order. Top: the prices (the variations of the average obtained price is in gray, the market VWAP in dark); bottom: the cumulated traded volume (grey) and market volume (dark).}
\label{fig:wimap}
\end{figure}

At time slice $t$, call $K_t \le K$ the number of currently active trading orders $\TO(k)$.
The $q \% $- quantile of the corresponding $K_t$ performance evaluations $\PE{}_t(k)$ is denoted by $\TH{}_t$. We consider $\TH{}_t$ as a $\PE{}$-threshold, separating ``bad trading performances'' (tagged ``1'') from ``normal trading performances'' (tagged ``0'').
We then \textit{binarize the performance evaluations $\PE{}_t(k)$} by setting 
\begin{equation}\label{def_Y}
\begin{split}
Y_t(k) = 1 \quad \text{if} \quad \PE{}_t(k) < \TH{}_t, \\
Y_t(k) = 0 \quad \text{if} \quad \PE{}_t(k) \geq \TH{}_t.
\end{split}
\end{equation}
In Figure \ref{fig:wimap}, we plot an example of the intraday behaviour of a trading algorithm. Its trading performance evaluation $\PE{}_t$ can observed within a trading day, in real-time, like some of the market context variable we used. A trading order may or may not be active at a given time slice as observed in Figure \ref{fig:perfheatmap}.

Figure \ref{fig:perfheatmapscore} displays synchronous intraday plots of trading performance evaluations in conjunction with the values of a few selected market variables. An essential goal of our methodology is, for each fixed time slice, to quantify on line the current influence of a market variable on trading performance degradation. Our automated online influence quantification replaces expert visual inspection of current trading orders performances, to identify critical market variables explaining trading performance degradations. For instance, visual inspection of Figures \ref{fig:perfheatmap} and \ref{fig:perfheatmapscore} will naturally ``explain'' the low performances observed at time slices $t = 39, 40, 41$ by the obvious trend changes simultaneously observed on rarity scores as well as by the volatility peak.

\begin{figure}[h!]
\centering
\includegraphics[width=\linewidth]{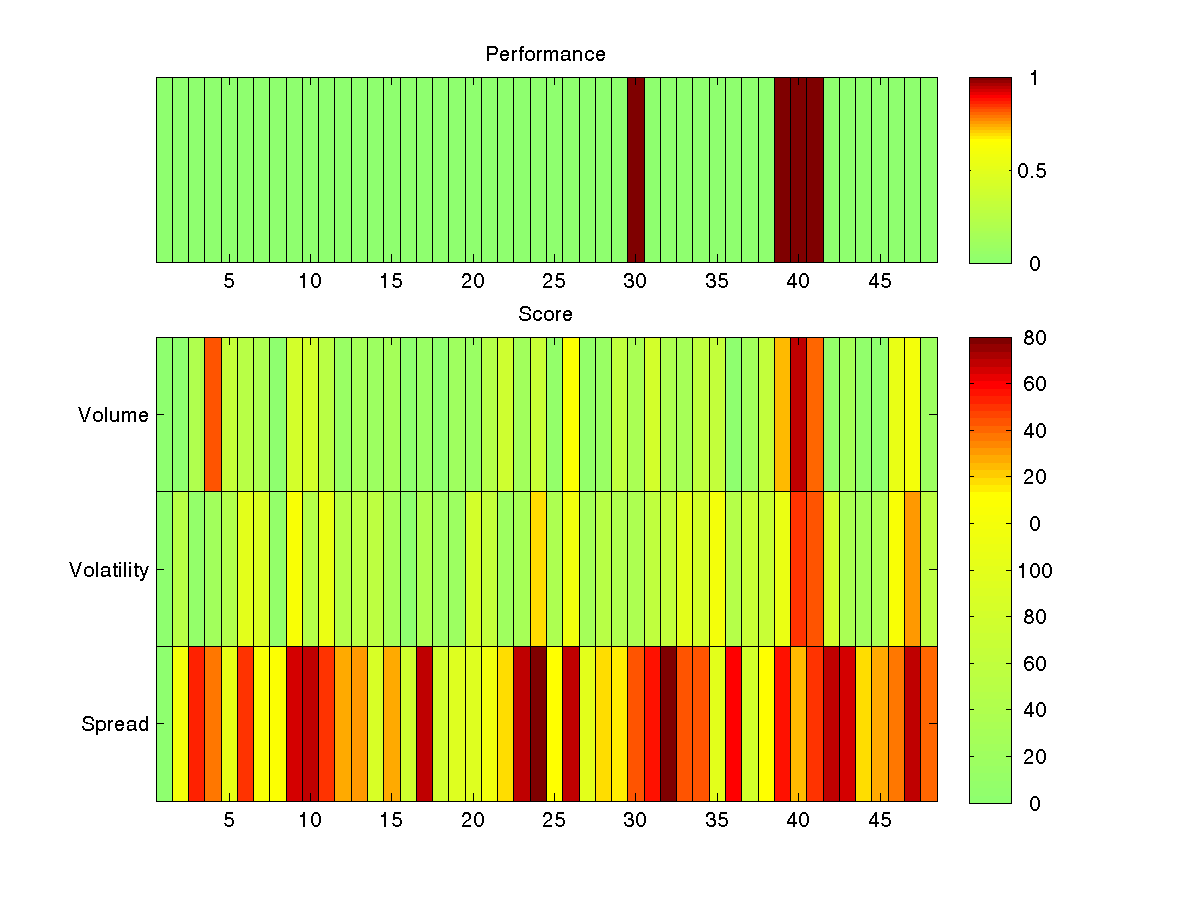}
\caption{Conjunction of the performance (top curve) of one traded order (first line of Figure \ref{fig:perfheatmapscore}) with the market context (bottom heatmap).}
\label{fig:perfheatmapscore}
\end{figure}

\section{Online anomaly detectors}
\label{detectors}

\subsection{Anomaly detection}

Online anomaly detection is a critical step in many applications, such as safety of complex systems, safety monitoring in automotive or aeronautics industries, remote health monitoring in biomedicine, real-time quality control for industrial production lines, etc
(See \cite{citeulike:10259244,citeulike:522801,citeulike:9302613,citeulike:522802} \cite{FISITA04}).\\
In the context of trading performance online monitoring, it is also quite natural to systematically enrich raw market descriptors by automated detection of anomalies affecting their dynamics. We have thus developed algorithmics dedicated to the online implementation of this processing step.

The occurrence of visually evident anomalies can be detected by algorithmic tracking of local regime changes in market descriptors dynamics, and may have potentially strong influence on performance degradation for the corresponding trading orders.
We have hence developed and implemented a set of 3 parametrized \emph{anomaly detectors}, dedicated to the online identification of ``significant'' \emph{Peaks or Crenels, Jumps, and Trend Changes} on generic time series. These 3 detectors automatically locate emerging anomalies, quantify their intensities, and filter them through adjustable gravity thresholds.

\subsection{Building online detectors}
\label{sec:detectors}

Consider a generic discrete time series $U_t$. A smoothed ``baseline'' $BU_t$ is generated as a moving local median of $U_t$. 
One then computes the local standard deviation $\sigma_t$ of the ``noise'' $U_t - BU_t$, and in turn, this defines ``outlier'' values of $U_t$.
Our three online anomaly detectors are based on local trend extractions at each time slice $t$ by fitting linear or quadratic regression models on short moving time windows to the left and the right of $t$. The detector parameters have simple geometric interpretations for the users and are kept fixed during online influence analysis.
Each anomaly detector is dedicated to a fixed type of anomaly, and generates a binary time series encoding the presence or absence of this anomaly type at successive time slices of $U_t$.

\begin{figure}[h!]
\centering
\includegraphics[width=\linewidth]{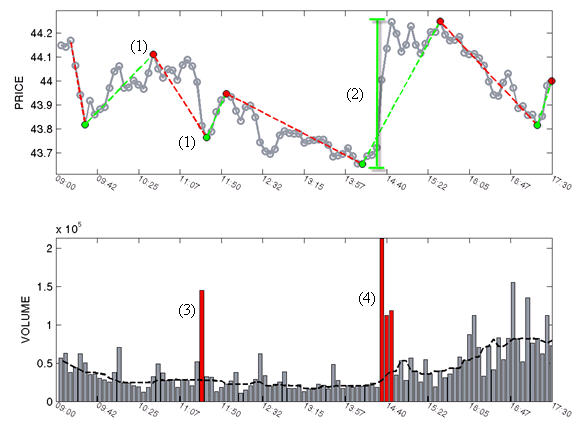}
\caption{The three abnormal patterns targeted by our three anomaly detectors: (1) price trends, (2) price jump, (3) volume peak, (4) volume crenel.}
\label{fig:detectorstylized}
\end{figure}

\subsubsection*{Peaks/ crenels detector}

A ``peak'' is the sudden occurrence of a high ``outlier'' value of $U_t$. 
More generally, a ``crenel'' is a cluster of successive high ``outliers'' with approximately equal values. 
Each crenel is described by 3 geometric ``crenel features'', namely, its \emph{time duration}, its \emph{thickness} (i.e., absolute difference between highest and lowest crenel points), and its \emph{height} above the baseline $BU_t$.
Minimal threshold values are imposed on these 3 features, as well as a minimal \textit{time gap} between successive crenels.

To detect peaks and/or crenels on the series $U_t$, one first extracts outliers with respect to the baseline $BU_t$; then one applies simple filters to detect local geometric configurations of outliers which satisfy the threshold constraints imposed on the three ``crenel features'' described above.
If a peak or crenel is detected at time $t$, then ``peak/crenel intensity'' $Peak_t$ is set equal to ``height'' of the peak/crenel above the baseline. If no peak or crenel is detected at time $t$, one sets $Peak_t = 0$.

\subsubsection*{Jumps detector}

A ``jump'' at time $t$ is a sudden level change between the $U_s$ values on finite time windows to the left and to the right of $t$. 
Bona fide jumps are described by 2 features, namely, a \textit{duration} $2L$ and a minimal \textit{jump size} $\Delta$. 
For each $t$, one fits two distinct quadratic regressions to the baseline $BU_s$, namely $Reg^{-}$ for $(t-1-L) \leq s \leq (t-1)$ and $Reg^{+}$ for $t \leq s \leq (t+L)$, where $L$ is a fixed parameter.

A jump is detected at $t$ if 
the ``jump size'' $JS(t)= |Reg^{+}(t) - Reg^{-}(t)|$ is larger than $ \Delta$, provided the two regressions have small enough residuals.
If a jump is detected at time $t$ on the series $U_t$, then ``jump intensity'' $Jump_t$ is set equal to the ``jump size'' $JS(t)$. If no trend change is detected at time $t$, one sets $Jump_t = 0$.

\subsubsection*{Trend changes detector}

Bona fide ``trend changes'' are described by 3 features, a \textit{duration} $2L$, a minimal \textit{slope change} $\lambda$, and a \textit{continuity modulus} $\varepsilon$.
For each $t$, one fits as above two quadratic regressions to the baseline $BU_s$, namely $Reg^{-}$ to the left of $t$ and $Reg^{+}$ to the right of $t$. 
Call $\alpha^{+} , \alpha^{-} $ the slopes of $Reg^{+}$, $ Reg^{-}$, and define the ``trend change size'' 
$$
TCS(t) = | \alpha^{+} - \alpha^{-} |.
$$
A local ``trend change'' is detected at time $t$ if $Reg^{+}, Reg^{-}$ have sufficiently small residuals and verify
$$
TCS(t) > \lambda \; ; \quad \text{and } \; |Reg^{+}(t) - Reg^{-}(t)| < \varepsilon.
$$
If a Trend Change is detected at time $t$ on the series $U_t$, then ``trend change intensity'' $Trend_t$ is set equal to the ``trend change size'' $TCS(t)$. If no jump is detected at time $t$, one sets $Trend_t = 0$.
%


\section{Probabilistic framework for influence analysis }

\subsection{The enriched set of explanatory factors}

For each trading order $\TO(k)$ and each market descriptor $M^j$, online analysis of the time series $t \rightarrow U_t = M^j_t(k)$ by the 3 anomaly detectors progressively generates 3 time series of ``anomaly intensities'' 
$Peak^j_t(k) , Jump^j_t(k) , Trend^j_t(k)$ which respectively encode the ``anomaly intensities'' of the Peaks/Crenels, Jumps, Trend Changes detected on the time series $U_t$. 
Applying the 3 anomaly detectors to our 7 basic market descriptors yields a set of 
$3 \times 7 = 21$ time series $A^1_t, \ldots, A^{21}_t $ of anomaly intensities.
Let $A_t = A^j_t$ be any one of these 21 anomaly intensities. A high value of $A_s$ detected at time $s$ may have a degradation influence on trading performances observed not only at time slice $s$ but also on performances observed at later time slices $t \in [ s , s + \tau ] $, where the short time lag $\tau$ is fixed (as a user selected parameter).
When we analyze below the influence at time $t$ of detected anomalies, we will hence take account of all the recent anomaly intensities $A_s$ where $ t- \tau \leq s \leq t$. To this end, we will replace each one our 21 anomaly intensities $A_t$ by a \emph{smoothed anomaly intensity at scale $\tau$} $[[A_t|\tau]]$ defined by the following formula,
$$ 
[[ A_t |\tau]] = \max_{t- \tau \leq s \leq t } A_s.
$$
From now on we will assume that a time scale $\tau$ is fixed, and note $[[ A_t]]$ instead of $[[ A_t |\tau]]$.
Note that $[[ A_t ]] \geq 0$ records the maximal gravity of recent anomalies (of fixed type) affecting the dynamics of a fixed market variable, and that for ``most'' time slices $t$ one has 
$ [[A _t ]] = 0$ for most scales $\tau$. 
We thus generate online $N=28$ time series $X^1_t ,\ldots , X^{N}_t$ of \textit{ explanatory factors}, namely the 7 current market descriptors $M^1_t , \ldots , M^7_t$ and their associated 21 smoothed anomaly intensities $ [[ A^1_t ]], \ldots, [[ A^{21}_t ]]$ at chosen time scales.

At each time slice $t$, and for each trading order $\TO(k)$, denote by 
$\X_t(k) \in \R^{N} $ the vector 
$$
\X_t(k) = \left[ \, X^1_t (k),\ldots , X^{N}_t (k) \, \right],
$$
which regroups the current values of our $N=28$ explanatory factors. 

In the online context, the $K \times N$ time series of explanatory factors become progressively available, and provide at each time slice $t$ the $K_t \times N$ incoming new values $ X^j_t(k) $, where $1\leq k \leq K_t \leq K, \; 1 \leq j \leq N$ and  $K_t$ is the number of trading algorithms handling  active orders at time $t$.
These  factors are viewed as potential ``causes'' for eventual degradations affecting the current binary performance evaluation $Y_t(k)$ of trading order $\TO(k)$ where $Y_t$ binarizes the performance evaluations $\PE{}$. The goal of our online influence analysis is to identify at each time $t$ the explanatory factors which have the most significant degradation influence on current trading performances of the whole portfolio.

\subsection{ Probabilistic framework }
\label{sec:perfQ}

To analyze  functional relationships between trading performance evaluations and explanatory factors, we introduce a time dependent  probabilistic framework.

Fix any  \emph{time slice} $t$.
We consider that each one of the $K_t$ trading order $\TO(k) $ currently active at time $t$ has been extracted at random from a very large finite pool $\Omega $ of ``virtually observable'' trading orders $\TO(\omega) $ , with $ \omega= 1, 2 , \ldots ,  \bar K$, where the fixed cardinal  $\bar K$ of $\Omega $ is much larger than  $ K_t$.

We consider $\Omega $ as a probability space endowed with the uniform probability.
The binary evaluations $Y_t(\omega)$ of current  trading performances defined in \eqref{def_Y} can then be viewed as a single  \textit{binary valued random variable} $Y_t$  defined on $\Omega$, verifying for each $\omega \in \Omega$
\begin{equation}
\begin{split}
Y_t(\omega) = 1 \quad \text{if} \quad \PE{}_t(\omega) < \TH{}_t, \\
Y_t(\omega) = 0 \quad \text{if} \quad \PE{}_t(\omega) \geq \TH{}_t.
\end{split}
\end{equation}
 Similarly all our $N$ explanatory factors $X^j_t$, $j = 1,2,\ldots, N$, can be viewed as real valued random variables $X^j_t(\omega)$ defined on $\Omega$ and for which we have only observed the $K_t$ values currently available at time $t$, namely the values $X^j_t(k) \in \R$ for $i=1, 2, \ldots , K_t $.
Then $\X_t = [ X^1_t, \ldots , X^N_t ]$ becomes a random vector defined  on $\Omega$, with values in $ \R^{N}$.
Our online influence analysis is performed anew at each fixed time slice $t$ and involves only the currently available $K_t$ joint observations of the random vector 
$\X_t $ and of  the random variable $Y_t  $. For each fixed $t$,  we will often  omit  the  subscript $t$ and adopt the abbreviated notations 
$$
Y= Y_t  \; ; \quad    X^j = X^j_t    \; ; \quad  \X=\X_t =  [ X^1, \ldots , X^N ] 
$$

\section{Binary valued predictors and their  predictive power }
\label{influence}

The time slice $t$ is kept fixed and deliberately omitted in this whole section, where we introduce  the precisely  relevant notions of  predictors, predictive power, and quantitative   influence of explanatory factors.

\subsection{Binary predictors } 
On a probability space $\Omega$ consider an arbitrary random vector $\X = X^1,  \ldots, X^N$ of  "explanatory factors" and an arbitrary  binary valued random variable $Y$. We call  \textit{binary valued  predictor} of the binary random variable $Y$ any random variable $\hat{Y}$ which is a deterministic function  of $\X$. Clearly $\hat{Y}$ is then  necessarily of the form $\hat{Y}_B = \one_B (X)$  where $\one_B $ is the indicator function of a fixed but  arbitrary Borel subset of  $\R^N \rightarrow \{0 ; 1\}$. \\
This class of  binary predictors is naturally imbedded in the convex set of "randomized" binary predictors $\hat{Y}_\phi$ of $Y$, indexed by arbitrary "decision functions" $\phi \in \Phi$, where $\Phi$ is the set of all Borel functions $\phi(x)$ defined for $x \in R^N$ and such that $0 \leq \phi(x) \leq 1$. The predictor $\hat{Y}_\phi $ defined by each  such $\phi$ verifies 
$$ 
P( \hat{Y}_\phi =1 | X ) = \phi(X)   \; ;  \quad \quad  P( \hat{Y}_\phi =0 | X) = 1 - \phi(X)
$$
Note that $\Phi$ is a closed compact convex subset of $ L_{\infty}(R^N)$, endowed with its  weak topology as dual of $ L_{1}(R^N)$.\\  

\subsection{Probabilities of correct predictions}

For randomized binary   predictors  $\hat{Y}_{\phi}$ of the true but yet unknown $Y$, the accuracy of  $\hat{Y}$  is usually characterized by the  two  conditional \emph{probabilities of correct prediction} $p^1$ and $p^0$, or equivalently by the two absolute probabilities of correct prediction   $P^1 = P^1(\phi)$ and $P^0 = P^0(\phi)$, defined by 
\begin{eqnarray}
\nonumber 
\label{eq:p1} p^1 &=& \Prob (\hat{Y}  = 1 | Y=1)  \quad \text{and } \quad  
P^1 = \Prob (\hat{Y}  = 1 ; Y=1)  \\
\nonumber 
\label{eq:p0} p^0 &=& \Prob (\hat{Y} = 0 | Y=0)  \quad \text{and } \quad  
P^0 = \Prob (\hat{Y}  = 0 ; Y=0) \\
\end{eqnarray}
The obvious expressions
$$
P^1  = \Esp (\phi(X) \one_{Y=1}) \; ; \quad \text{and }  P^0 = \Esp ( (1 - \phi(X) ) \one_{Y=0})
$$
show that $ P^1(\phi)$ and $ P^0(\phi)$ are weakly continuous functionals of $\phi \in  \Phi$. 
Intuitively the predictive power of a predictor should  be an increasing functional of $p^1$ and  $p^0 $, or equivalently  an increasing functional of $P^1$ and  $P^0 $. Indeed the classical $2 \times 2$ \textit{confusion matrix} for  the estimation of $Y$ by the binary predictor  $\hat{Y}$ is determined by $p^1 , p^0 $ as follows:
\begin{center}
\begin{tabular}{c | c | c |}
\multicolumn{1}{r}{}
& \multicolumn{1}{c}{$Y= 0$} 
& \multicolumn{1}{c}{$Y = 1$}  \\ 
\cline{2-3}
$\hat {Y}$ = 0 & $p^0$ & $1- p^1 $\\ 
\cline{2-3} 
$\hat {Y}$ = 1 & $1- p^0$ & $p^1$ \\ 
\cline{2-3}
\end{tabular}
\end{center}
\begin{figure}[h!]
\centering
\caption{Confusion Table for the estimation  of $Y$ by a binary predictor $\hat{Y}$}
\label{fig:condY}
\end{figure}
This motivates the following definition of predictive power.
\subsection{Predictive power}

For any random  vector of "explanatory variables" $X \in R^N$ and any binary random variable $Y$  jointly defined on a  probability space $\Omega$, the joint  probability distribution $\mu$ of $(X,Y)$ belongs to the compact convex set $\mathcal{M}$ of all probabilities on    $R^N \times \{0 ; 1\}$. 
Select and fix any non-negative continuous function $Q(\mu, a, b)$   of $\mu \in \mathcal{M}$  , $ 0 \leq a \leq 1 , 0 \leq b \leq 1 $ which is a \emph{separately increasing} function of  $a$ and $b$.   

We then define the  \emph{ predictive power  } $\pi(\phi) = \pi(\mu,\phi)$  of  each randomized binary predictors $\hat{Y}_{\phi}$ of $Y$ by
\begin{equation}
  \label{eq:predpower}
  \pi(\phi) = Q(\mu, P^1, P^0)  = Q(\mu, \mu^1 p^1 , \mu^0 p^0)  
\end{equation}
where $\mu^1 = \Prob(Y=1) $ and $ mu^0 = \Prob(Y=0)$.
Note that  in our benchmark study below, due to our adaptive binarization of trading performance,  the probabilities  $\mu^1 $ and $\mu^0 $  will be constant in time and will  have known pre-assigned fixed values such as $3\%$ and $97\%$.\\
The predictive power $\pi(\mu,\phi) $ is then  clearly continuous in $(\mu,\phi)$ for the weak convergence topologies of $\mathcal{M} $ and $\Phi$.

We shall see in the next paragraph that this definition of predictive power is compatible, and actually extends these definitions actually extend the predictive power quantification by relative entropy.\\ 
The functional $Q(\mu, P^1, P^0)$ will be called a \textit{predictor quality} function. Here are basic examples of functions $Q$ often used in the accuracy analysis of predictors.
\begin{itemize}
\item $Q_{min} = \min \{ P^1 , P^0 \}$,
\item $Q_{weight} = u\,   P^1 + (1-u) P^0$, for some $0 < u(\mu) < 1$,
\end{itemize}
\subsection{Predictive power based on  mutual information and/or  relative entropy} 
\label{sec:MIR}

An information theoretic characterization for  the predictive power  of a  predictor $Z= \hat{Y}$ of $Y$ is   the amount of information which $Z$ reveals on the yet unknown variable $Y$. This has classically been quantified by \emph{relative entropy}  criteria such as the mutual information ratio $\MIR(Z,Y)$  (see for instance \cite{brillinger2004some}\cite{azencott4analysis} \cite{mougeot2011traffic} \cite{billingsley1965ergodic} \cite{khinchin1957mathematical}\cite{shannon2001mathematical}). 
Recall that the entropy $H(U)$ of a random variable $U$ taking only a finite number of values $u_i$ is given by
\begin{equation*}
H(U) =  - \sum_{i} \Prob(U= u_i) \log \Prob(U = u_i),
\end{equation*}
The \emph{mutual information ratio}  $\MIR(Z,Y)$ between $Y$ and its  predictor 
$ Z= \hat{Y}$ is  defined by
$$
\MIR(Z,Y) = \frac{H(Z)+H(Y) - H(Z,Y)}{H(Y)},
$$
where $H(Z), H(Y), H(Z,Y)$ are the respective entropies of the three random  variables $Z, Y$ and  $(Z,Y)$. Here $Z$ and $Y$ are binary valued and $(Z,Y)$ takes only 4 values. 

The ratio $\MIR(Z,Y)$  which is directly related to the relative entropy of $Z$ with respect to $Y$ lies between 0 and 1,  and reaches the value 1 if and only if Y is a deterministic function of Z. Good predictors $Z$ of $Y$ should  thus achieve  high values of $MIR(Z,Y)$. Indeed we have the following result .
\begin{proposition} 
Fix any random  vector of "explanatory variables" $X \in \R^N$ and any binary random variable $Y$  jointly defined on a  probability space $\Omega$, and call $\mu$ the joint  probability distribution  of $(X,Y)$. Assume that $Y$ is not deterministic, so that $H(Y) > 0$. Then for all randomized binary predictor $Z = \hat{Y}_{\phi}$ of $Y$  defined by arbitrary  Borel decision functions $0 \leq \phi \leq 1$, the mutual information ratio $\MIR( Z,Y)$ is a separately increasing functional  of the two conditional  probabilities of   correct decisions $ p^1 = \Prob (\hat{Y}  = 1 | Y=1) $  and $ p^0 = \Prob (\hat{Y}= 0 | Y=0)$, provided  $ p^1 \geq 1/2 $ and $ p^0  \geq 1/2 $.
\end{proposition}
Proof:  The joint distribution of $(Z,Y)$ on $\{ 0 ; 1 \}^2$ is easily seen to be determined by the 3 parameters $p^1, p^0$, and $ \mu^1 = \Prob(Y=1)$. Since  $H(Y)$ is fixed,  $\MIR( Z,Y)$ is an increasing linear function of $H(Z) - H(Z,Y)$. An elementary computation easily proves the identity 
$$
H(Z) - H(Z,Y) =  - \mu^1 Ent(p^1) - \mu^0 Ent(p^0)
$$
where $ Ent(p) = - p \log (p) - (1-p) \log(1-p)$ and $\mu^0 = 1- \mu^1$. Since the entropy $Ent(p)$ decreases  with $p$ for $p \geq 1/2$ , this concludes the proof.

%

\subsection{Generic optimal randomized  predictors } 
\label{generic}

We now characterize the randomized predictors achieving optimal predictive power..
\begin{proposition}\label{optimal.predict}
Fix a random vector $\X \in R^N$ of explanatory factors and a target binary variable $Y$.  Let 
$0 \leq v(x) \leq 1 $ be any Borel function of $x \in R^N$ such that $v(\X) = \Prob(Y= 1 \; | \; X)$ almost surely.

For any Borel decision function $\phi \in \Phi$, define the predictive power of the  randomized predictor $\hat{Y}_{\phi}$ by  $\pi(\phi) = Q(\mu, P^1(\phi), P^0(\phi))  $, where $Q$ is a fixed continuous and  increasing function of the probabilities of correct decisions $P^1, P^0$ .
Then there exists  $\psi \in \Phi $  such that the  predictor $\hat{Y}_{\phi}$  has maximum predictive power  
$$
\pi(\psi) = \max_{\phi \in \Phi } \pi(\phi)
$$
Any such optimal Borel function $0 \leq \psi(x) \leq 1$  must    almost surely verify,  for some suitably selected constant $0 \leq c \leq 1$.
\begin{equation} \label{opt}
\psi(X) = 1 \; \text{for }  v(X)  > c \; ; \quad  \psi(X) = 0 \; \text{for }  v(X) < c
\end{equation}
\end{proposition}
Proof: Predicting the actual value of the yet "unknown" binary random variable $Y$ given the random vector $\X$ of explanatory variables  is clearly equivalent to deciding between the two formal ``hypotheses'':
$$
\Hyp^0 : \{Y= 0 \} \text{ versus } \Hyp^1 : \{ Y_t = 1 \}
$$
on the basis of the observed  $\X$, which is a standard  \textit{testing problem}  \cite{citeulike:1506124}. Any borelian ``rejection region'' $B \subset \R^k$ defines the  binary valued  predictor $\one_B(YX $ of $Y$ which  rejects $\Hyp^0$ whenever $\X \in B$. More generally any Borel decision  function $0 \leq \phi(x) \leq 1 $ defines the binary predictor $\hat{Y}_{\phi}$ which, given the observed $\X$ rejects $\Hyp^0$ with probability $\phi(X)$. This test has ``confidence level'' $\alpha = 1- P^0 $, and ``detection power'' $ P^1$ where $P^1, P^0$ are the probabilities of correct decisions for the predictor ${\hat Y}_{\phi}$.

Classical  testing of $\Hyp^0$ versus $\Hyp^1 $ involves the   \emph{likelihood function} defined with probability 1 by
$$
L(\X)  = \frac{  \Prob ( Y = 1 \, | \, \X ) } { \Prob ( Y = 0 \, | \, \X) } = \frac{v(X)}{1-v(X)}   
$$
By  Neymann-Pearson theorem ( \cite{citeulike:1506124}, for each "confidence level" 
$0 \leq  (1- \alpha) \leq 1$ there exists a randomized binary predictor $\hat{Y}_{\phi}$ which  maximizes $P^1(\phi)$ among all predictors verifying $P^0(\phi) \geq 1- \alpha$. Morover one can find $c \geq 0 $ such that the Borel function $0 \leq \psi(x) \leq 1$ verifies almost surely
\begin{equation} \label{psi}
\psi(X) = 1 \; \text{for }  L(X) > c \; ; \quad  \psi(X) = 0 \; \text{for }  L(X) < c
\end{equation}
Since both $P^0 = P^0 (\phi)$ and $P^1 = P^1 (\phi)$ are weakly continuous  functions of  $\phi \in \Phi $,  the predictive power $\pi(\phi) = Q(\mu, P^1(\phi), P^0 (\phi))$ of $\hat{Y}_{\phi}$ is also weakly continuous in $\phi$ and thus must reach its maximum on the weakly compact set $\Phi$ for some Borel function  $\phi^* \in \Phi$. 

Since $ Q(\mu, P^1, P^0)$ is an increasing function of $P^1 $ and $P^0$, we see that the optimal predictor $\hat{Y}_{\phi^*}$ must necessarily maximize $P^1(\phi)$ among all predictors verifying $P^0(\phi) \geq P^0(\phi^*)$. Select the confidence level $1- \alpha = P^0(\phi^*)$  and apply the Neyman-Pearson theorem just recalled above to conclude that  there exists a threshold $c \geq 0 $  and an associated  $\psi $ of the form \eqref{psi}such that $P^0(\psi) \geq P^0(\phi^*)$ and $P^1(\psi) \geq P^1(\phi^*)$. This implies the following inequality between predictive powers
$$ 
\pi(\psi) = Q( \mu, P^1(\psi) , P^0(\psi) ) \geq Q( \mu, P^1(\phi^*) , P^0(\phi^*) ) = \pi(\phi^*)   
$$
and hence $\pi(\psi) = \pi(\phi^*)$ sice $\hat{Y}_{\phi^*}$ has maximal predictive power. This clearly achieves the proof.
\begin{definition} \label{influence}
In the preceding situation we will quantify  the capacity of the random vector   $\X$ to "explain" the target binary variable $Y$ by an "influence coefficient" $\mathcal{I}(X,Y)$ defined as the predictive power $\pi(\psi)$ of an optimal randomized binary predictor  $\hat{Y}_{\psi}$ of $Y$. More precisely the influence coefficient of $\X$ on $Y$ is given by 
$$
\mathcal{I}(X,Y) =  \pi(\psi) = \max_{\phi \in \Phi } \pi(\phi)
$$ 
\end{definition}
Clearly, once the quality function $Q$ is selected and fixed, the influence $\mathcal{I}(X,Y) $ depends only on the joint probability distribution $\mu$ of $(X,Y)$.

The notion of influence coefficient is immediately extended to arbitrary subsets of explanatory factors.  To any subset of indices $G \subset \{1, 2, \ldots , N\}$, we associate the  random vector  $X^G= \{ X^{j}  \;| \; j \in G \}$ of explanatory factors, with $\Card(G) \leq N$, and we define as above the influence coefficient by
\begin{equation} \label{J(G)}
J(G) = \mathcal{I}(X^G ,Y)  \leq \mathcal{I}(X,Y) 
\end{equation}

\section{Quantifying on line the influence of  groups of explanatory factors}
\subsection{Benchmark study context}

In our intraday data  study below, at each fixed time slice $t$, we observe simultaneously on $K_t <700 $ trading lines the current values of our random vector $\X \in R^N$ of $N = 28$ explanatory factors (see section \ref{detectors}), and the corresponding current values of the binarized  trading performances $Y$. Ideally, for each subgroup $X^G= \{ X^{j}  \;| \; j \in G \}$ of explanatory factors, where  $G \subset \{1, 2, \ldots , N\}$, we want to estimate  the current  \textit{influence coefficient} $J(G)$ of $X^G$ on $Y$  by the formula \eqref{J(G)}, using only the current  sample of $K_t$  jointly observed values of $(X^G, Y)$. Statistical reliability of the $J(G)$ estimates will lead us below to consider only groups $G$ of small cardinal.  

Our goal was to determine, at each time slice $t$, one or possibly several groups $G$ of explanatory factors having small cardinal and high  influence $J(G)$ on current trading performances, and to specifically focus on detecting    small groups of  current ``major  causes'' for  the trading performance \emph{degradations} just  observed at time $t$. 

To this end we selected  a  class of asymmetric  predictive power functionals parametrized by one parameter $ 70 \% \leq r \leq 100 \%$, called here  the ``floor predictive power''.  For each predictor $\hat{Y}$ of $Y$ with current conditional probabilities of correct predictions $p^1, p^0$, the predictive power of $\hat{Y}$ was computed by  
\begin{equation}\label{quality}
\begin{array}{lclcl}
Q_r ( p^1, p^0) &=& p^1, & \text {if} & \min(p^1 , p^0) \geq  r \%,\\
Q_r ( p^1, p^0) &=& 0 , & \text {if} & \min(p^1 , p^0) < r \%.
\end{array}
\end{equation}
Note that $Q_r$ emphasizes strongly the probability $p^1$ of correctly predicting bad trading performances. The associated influence coefficients $J(G) $  then quantify the current  impact  of the explanatory factors  $X^G$ on ``performance degradation''. In our benchmark study of intraday datasets, systematic tests led us to fix  $ r= 85\%$. 

\subsection{Influence computation: accuracy analysis}
\label{compute.influence}

As in the preceding subsection, the time slice  $t$ is fixed and we keep the same notations. We now analyze  how to implement a  numerical computation of the current influence coefficients $J(G)$  for small groups $X^G$ of explanatory factors. Let $m $ be the cardinal of $G$ and denote  $X^G = Z = [ \;  Z^1, \ldots, Z^m  \; ] $. To compute $J(G)$ we need to compute an optimal  decision function $0 \leq \psi(Z) \leq 1$ , maximizing the  predicting power of the predictor $\hat {Y} $ defined by $\Prob \{ \hat{Y} = 1 |  Z \} = \psi(Z)$. By proposition  \ref{optimal.predict} , for  each value $z$ of $Z$ currently  observed at time $t$, this requires first to estimate  by empirical frequencies $\hat{v}(z)$ the  probabilities 
$$ 
v(z) = \Prob \, [ \, (Z=z ) \cap (Y = 1) \, ] 
$$
and then to find an optimal threshold $0 < c  <1 $ for the $\hat{v}(z)$ values. 
At time $t$ the estimates $\hat{v}(z)$ are derived only from the  moderately sized sample of $K_t \equiv  700$  of currently  observed joint values for the pair $(Z,Y)$. By construction of the binarized trading performance  (see section \ref{input}) the empirical frequency $\{ Y=1 \}$ is kept constant equal to $3\% \leq q \leq 5\% $. So the number of currently observed  values $z$ of  $Z$ for which $\hat{v}(z) $ is non zero will always be inferior to $K_t \times 5/100 \simeq 35$. Empirical thresholding of the  $\hat{v}(z)$ at time $t$ can then obviously be restricted  to exploring at most 35 values of $c$.

 We seek then an  optimal  decision function $0  \leq \psi(Z) \leq 1$, which according to formula \eqref{opt}, should be associated to  some   threshold $1 > c >0$, with $\psi(Z) =1 $ when $v(Z) > c$ and $\psi(Z) =0 $ when $v(Z) < c$. To achieve statistical robustness and fast online computation,  we  restrict    $\psi(Z)$ to only take the values $0$ or $1$, and we thus  impose  $\psi(Z) = \one_{v(Z) \geq c}$.  At time $t$, the predictive power $g(c) = \pi(\psi) = Q(P^1,P^0)$ of the  estimator $\hat{Y} = \psi(Z)$ depends only on its current  probabilities of correct prediction 
$$
P^1 = \Prob \{ (v(Z) > c ) \cap (Y=1) \} \qquad P^0 = \Prob \{ (v(Z) < c ) \cap (Y=0) \}  
$$ 
At time $t$,  these probabilities  are   approximated by current empirical frequencies 
$\hat{P}^1,  \hat{P}^0$ after replacing $v(Z)$ by the current estimates $\hat{v}(Z)$. This yields an estimated predictive power $\hat{g}(c)$ and the optimal threshold $c^*$  is selected from at most 35 threshold values   by a trivial  maximization of $\hat{g}(c)$. The current estimate of the influence coefficient of the group of factors $G$ is then $\hat{J}(G) = \hat{g}(c^*)$.

The practical accuracy of the estimates $\hat{J}(G)$ is strongly determined by the accuracy of the estimators $\hat{v}(z)$ of the probability $v(z) $. Given the moderate size of $K_t$ , we deliberately \emph{discretize} all our  explanatory factors so that the vector $Z= X^G$ is restricted to take only a finite number $s$ of distinct values. Under the favourable assumption of approximate independence of the $K_t$ trading orders observed at time $t$ the errors of estimations $\varepsilon(z) =  \hat{v}(z) - v(z) $ have standard deviations  
${  v(z) (1 - v(z) ) } /{ K_t }$ and hence the relative errors $\varepsilon(z) / v(z)$ on $v(z)$ are of the order of ${1}/ {\sqrt{v(z) K_t} }$, which is inferior to  
${1} /{\sqrt{w  K_t} }$ where $w = \min_{z} v(z)$.

The optimal threshold $c^*$ is computed above by empirically thresholding a set of at most 35 estimated $\hat{v}(z)$, since 
$5\% \times K_t$ is of the order of at most 35 in our data set. The relative error of estimation on $c^*$ will then be roughly of the order of ${1}/ {\sqrt{w  K_t} }$.
Thus to obtain a relative error inferior to, say, $11 \%$ for the estimation of the optimal $c^*$, one needs to at least impose the constraint $\frac{1}{ w K_t } \; < \; 1.21 /100$, wich yields 
$ 1/w < \frac{1.21 K_t}{100}$. Since  $\Prob(Z=z) \geq  w $ for each $z$ in the currently  observed range of $Z = X^G$ which after discretization contains $s$ values,  we have $ 1 \geq s w$ and hence   $ s < 1/w <  \frac{1.21 K_t}{100}$.  In our benchmark intraday data, the number $K_t$  of orders active at time $t$  is $K_t \equiv 700$ , and  hence  the cardinal of observable values for the discretized  random vector $Z = X^G$ should be less than 8.


Maximal discretization of each explanatory factor is reached when each factor is binarized. But even in this case, since the vector $Z = X^G$ is of dimension $m = $ cardinal(G), the number $s$ of distinct values for $X^G$ is $s = 2^m$ and hence we must still impose the constraint cardinal(G)  $  \leq 3$. 

These cardinality constraints show that:
\begin{itemize}
\item for cardinal($G$) equal to  3, all three factors  in $X^G$ must be binarized;
\item for cardinal($G$) equal to  2, one of the two  factors  in $X^G$ must be binarized and the other one must be discretized with at most three values;
\item for cardinal($G$) equal to 1 , the single factor involved must be discretized with at most 8 values.
\end{itemize}
In our benchmark study of intraday data the number $ K_t $ of  trading orders concretely ranged anywhere  between 400 and 700. The preceding analysis thus showed that at each time $t$ and for any group $G$ of explanatory factors,  the computation of the influence coefficient $J(G)$ could only be statistically robust if cardinal($G$) was equal to 1 or 2 and if  each explanatory factor was  binarized. 
In practice, at each time $t$,  a key computing  task is to select an optimal binarization of each explanatory factor, as indicated below.
\section{Influence computation: optimized discretization of explanatory factors } \label{discretization}
\subsection{Influence computation for  a single explanatory factor}\label{singlefactor}

Consider any real valued single explanatory factor $Z$ having a continuous conditional density function $w(z)$  given $Y=1$.  To predict $Y$ given $Z$ the best binarization of $Z$ should select  a  subset $B$ of $S$ maximizing the predictive power of the predictor  $\one_B (Z) $. An easy extension of the proposition \ref{optimal.predict} proved above, shows that  any optimal  $B$ should be the set of all $z \in \R$ such that $ w(z) > c $ for some $c >0$. Thus an optimal $B$ must be   a closed  level set $L$ of the unknown conditional density function $w(z)$. The family of all closed  sets in $\R$ is well known to have infinite Vapnik-Cervonenkis dimension (see \cite{vapnik1971uniform},\cite{vapnik1971uniform},\cite{vapnik2000nature}). So, in view of Vapnik's theorems on  automated learning  (see \cite{vapnik2006estimation},\cite{cristianini2000introduction}), empirical optimal selection  of $B $ among all closed sets   will have weak  generalization capacity, increasing extremely slowly with  the number  $K_t$ of data.  This has  naturally led us to select sub-optimal but much more robust classes of predictors, whith radically reduced  Vapnik-Cervonenkis dimension. \\
In the cases where   $w(z)$ can  be considered as roughly  unimodal or monotonous, the level sets of $w$ are unions of at most two disjoint intervals.   We  thus deliberately restrict our class of binary predictors of $Y$ to two-sided ones:
\begin{defin}[Two-sided binary predictor]\label{def:2pred}
  \emph{Two-sided binary predictors} are of the form $h_{\theta} = \one_B(Z)$ where $B$ is the union of the two disjoint intervals $(-\infty, \theta^{-})$ and  $ (\theta^{+}, +\infty)$ , indexed by the vector  $\theta= (\theta^{-} , \theta^{+}) \in \R^2$, with  $\theta^{-} < \theta^{+}$.
\end{defin}

\begin{figure}[h!]
\centering
\includegraphics[width=.8\textwidth]{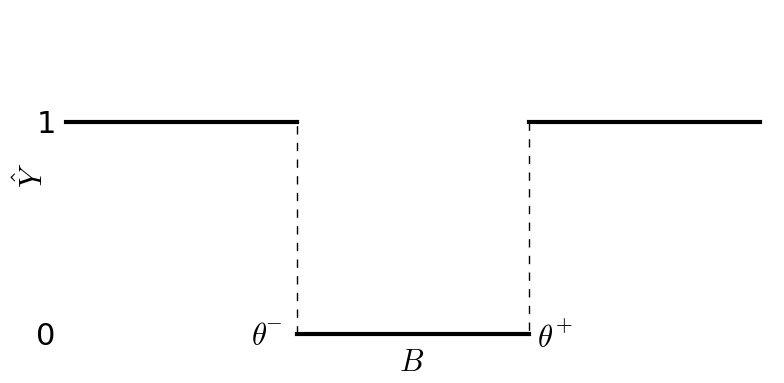}
\caption{Typical  predictor defined by  two intervals $(-\infty, \theta^{-}) $ and  $(\theta^{+}, +\infty)$.}
\label{fig:BorelExample}
\end{figure}
Note that $h_{\theta} $ predicts bad trading performances if and only if the explanatory factor $Z$ takes sufficiently large or sufficiently small values. Hence these estimators of trading performance degradation have \emph{an immediate interpretability} for natural users of online trading performance monitoring.

At time $t$, given the current $K_t$ joint observations of the explanatory factor $Z$ and of the binarized trading performance $Y$, an immediate counting provides for each $\theta $ the empirical estimates $\hat{P}^1$ and  $\hat{P}^0$ of the probabilities of correct prediction  
$P^1, P^0$ for the estimator $h_{\theta}$, given by 
$$
P^1 = \Prob\{ (Z < \theta^{-}) \cap (Y=1) \} + \Prob\{ (Z > \theta^{+}) \cap (Y=1) \}
$$
$$
P^0 = \Prob\{ ( \theta^{-} <  Z < \theta^{+}) \cap (Y=0) \} 
$$
The predictive power $\pi(\theta)$ of $h_{\theta}$ is then readily  estimated by the explicit formula 
$$
\hat{\pi}(\theta) = Q_r ( \hat{P}^1 , \hat{P}^0).
$$
The influence coefficient $\mathcal{I}(Z,Y)$  of $Z$ on $Y$ at time $t$ is then estimated by  maximizing  $\hat{\pi}(\theta)$ over all $\theta$ in $\R^2$. At time $t$, the set $S$ of  currently observed values of  Z has  cardinal  inferior or equal to    $K_t$. The previous formulas show that to  maximize $\hat{\pi}(\theta)$,  we may in fact  restrict both $\theta^{-}$ and $ \theta^{+}$ to belong to $S$, so one needs only explore at most $K_t^2  / 2 $ values of $\theta$. 

Clearly this  computation tends to   underestimate the influence $\mathcal{I}(Z,Y) $ . Nevertheless in our benchmark studies we have systematically  applied this approach for the following  reasons.
\begin{itemize}
\item the set of binary predictors $h_{\theta}$ has the merit of having { finite Vapnik-Cervonenkis dimension equal to 2} , so that our empirical estimate of the maximum of $\pi(\theta)$ will be statistically robust even for  moderate realistic values of $K_t \equiv 700$;
\item the immediate interpretability of the predictors $h_{\theta}$  enables user friendly online graphic displays of the explanatory factors currently having  high  influence on performance degradation.
\item at each time $t$, at most $K_t^3$ basic operations suffice to implement the brute force maximization of $\pi (\theta)$ which generates our current evaluation of $\mathcal{I}(Z, Y)$.
\end{itemize} 
Note that the two optimal thresholds $(\theta^{-} , \theta^{+})$  will of course strongly depend on the time slice $t$. 

When the explanatory factor $Z$ is one of the 21 \emph{smoothed anomaly detectors } $[[ A_t ]] \geq 0$ introduced above in section \ref{detectors}, the preceding implementation can be simplified. Recall that $[[ A_t ]]$ records the maximal gravity of very recent anomalies of fixed type affecting the dynamics of a fixed market variable, and that for ``most'' time slices, $[[ A_t ]]$ takes the values $0$. Thus it is natural to expect that only higher values of $[[ A_t ]]$ to be potential explanations for currently degraded trading performances.
So for practical applications to $Z= [[ A_t ]]$ of the preceding approach, we may actually impose the constraint $\theta^{-} = 0$, with essentially no loss of predictive power.

\subsection{A few examples for single explanatory factors}

The empirical strategy just presented to estimate the influence $J(G)$ when cardinal(G) equals 1 has been numerically validated on our benchmark set of intraday data. We now outline a few examples.
Recall that our benchmark study used the predictive power functional $\pi = Q_r(P^1, P^0)$ given above by formula \eqref{eq:predpower}, which  is specifically sensitive to predictors capacity to detect\emph{ degradations} of  trading performances. On  our  intraday data sets,   we have methodically tested the values $r = 70 \% , 75 \% , 80 \% , 85 \% , 90 \% , 95 \%$ for the "floor predictive power" $r$ ; the value  $r= 85\%$ turned out to be  the best choice for these data sets, and was adopted for all results presented below.

Figure \ref{imp} illustrates for the fixed time slice $t = 45$, the predictive power of the predictors $h_{\theta}$ based on the single market variable $Z =$ ``Momentum in Bid-Ask Spread'' .

The trading performance evaluator $\PE{}$ is the ``slippage in bid-ask spread''. The $\PE{}$-thresholds $\TH{}_t = \TH{}_{45}$ determining low trading performance is fixed at the 3\% -quantile of all performance evaluations observed at time $t=45$. \\
The $x$ and $y$ axes in the graph (figure \ref{imp}) indicate the threshold values  $(\theta^{-}, \theta^{+})$ for the market variable $Z =$ ``Momentum in Bid-Ask Spread''. The $z$ axis displays  the predictive power $\pi (\theta) $ of $h_{\theta}$. The red marker indicates at time $t= 45$, the estimated influence coefficient  $\mathcal{I}(Z,Y)$ on $Y$ for  this specific  market variable $Z$, which turns out to be   equal to  $ 100\%$. The threshold vector $\theta  = \theta_{45}$ which achieves maximum predictive power at time 45 is equal to  
$( 66.76 , \; 3.87 ) $.
At each time $t$, our 7 basic market variables can then be ranked on the basis of their approximate influence values computed as above, which provides a ranking of their respective capacity to explain current bad trading performances.
\begin{figure}[h!]
\centering
\includegraphics[width=\linewidth]{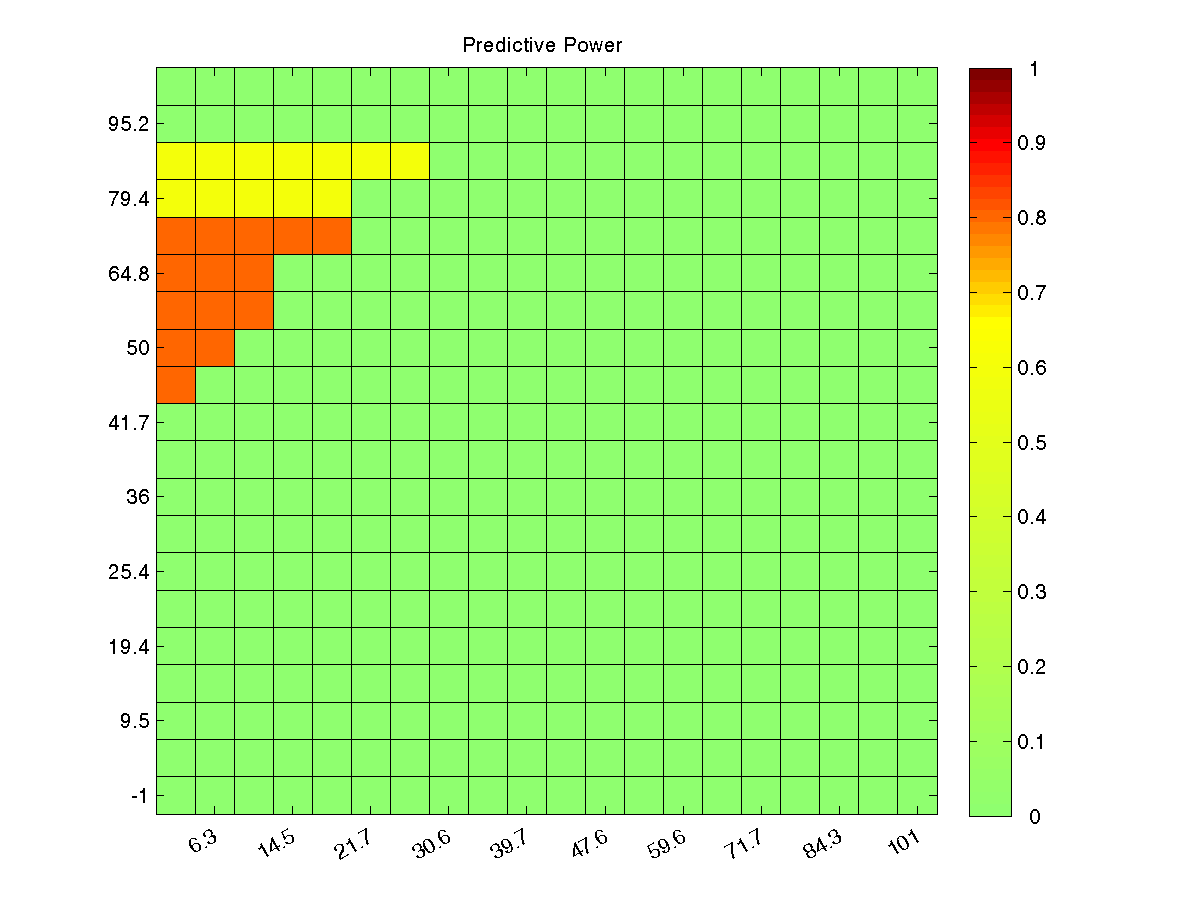}
\caption{Predictive power as a functional of the two thresholds $\theta^+$ (y-axis) and $\theta^-$ (x-axis) of the market variable \emph{Volume Score} at time slice $t=72$. It can be seen that $\theta^-$ lower than 10 and $\theta^+$ around 65 generate an efficient predictor of bad trading performance during this time slice.}
\label{imp}
\end{figure}
\section{Influence computation for pairs of explanatory factors}
\label{twofactors}

Again at fixed time $t$, we now sketch our online ``optimized fusion'' of predictors to estimate  the influence coefficient $J(G)$ when $G$ is a group of 2 explanatory factors 
$Z = [ \; Z^1, Z^2 \; ]$. Our  statistical robustness analysis above indicates the necessity to consider only classes  of trading performance predictors having radically  low Vapnik-Cervonenkis dimension. So our predictive power maximization among predictors based on $Z$ is deliberately restricted to   the following class  of predictors. 

Let  $ \POL_2$  be the set of all 16 functions mapping $\{0 ; 1\}^2$ into 
$\{ 0 ; 1 \}$. The class  $\mathcal{H}$ will be the set of all predictors of the form  
$$
h (Z) = \mp ( f(Z_1), g(Z_2) ) 
$$
where   $\mp \in \POL_2$, and  the  indicator functions $f(z)$ and $g(z)$ are both two-sided binary predictors in the sense of definition \ref{def:2pred}. The class of  binary predictors $\mathcal{H}$ has  Vapnik-Cervonenkis dimension equal to 4. Hence the estimation of maximal predictive power within $\mathcal{H}$ by empirical estimation of probabilities $P^1$ and $P^0$ on the basis of the current $K_t$ joint observations of $Y, Z_1,  Z_2$ will be statistically robust. This provides at time $t$ a  stable estimator of $J(G)$, which as above tends to undervalue the true $J(G)$.

In concrete implementation of this approach at fixed time $t$, we first select only  pairs of predictors $f(Z_1), g(Z_2)$ which already have reasonably high  probabilities of correctly predicting $Y$. To maximize the predictive power of $h(Z) = \mp ( f(Z_1), g(Z_2) )$ we need to select the best  binary polynomial $\mp$ among the 16 elements of $\POL_2$. We then impose  $\mp(0,0) = 0 $ and $\mp(1,1) = 1$, so that whenever the predictions of $f(Z_1), $ and $g(Z_2)$ agree, we also have $h(Z) = f(Z_1) =  g(Z_2)$. This "accelerated fusion"  is fairly classical in multi-experts fusion (see \cite{alkoot1999experimental}) and obviously provides an acceleration multiplier  of 4 in the online computation of $J(G)$. 

For groups $G$ of $k=3$ or $k=4$ explanatory factors,  one could estimate $J(G)$ by similar  sub-optimal but implementable strategies. However  the corresponding predictor classes have Vapnik-Cervonenkis dimensions 6 and 8, and their statistical robustness is hence much weaker  in the concrete context of  our intraday datas set, since  at each time $t$ the key Vapnik ratios $K_t/6$  and $K_t/8$  were resp. inferior to 120 and 70, values which are much too small and strongly suggested  to avoid the estimation of $J(G)$ for cardinal(G) $\geq 3$. 

\section{Numerical results}
\label{output}

We now  present the numerical results obtained by applying the above methodology to our benchmark dataset of intra-day trading records.

\subsection{Dataset (portfolio) description}

Recall that our intra-day benchmark data involve a total of 79 time slices of 5 minutes each (i.e. this portfolio has been traded from 8:55 to 15:30 London time), and that we are monitoring a portfolio of 1037 trading orders, with a maximum of $700$ trading orders active simultaneously at each time slice. 

At each fixed  time slice $t$, we compute the current influence coefficient for each one of our 28 explanatory factors, namely the 7 market descriptors $M^j$ themselves and the $3 \times 7$ smoothed anomaly detectors monitoring the dynamic of these market descriptors.

These 28 explanatory factors generate $378 = 28 \times 27 / 2 $ pairs of factors.  By accelerated  fusion as above, we compute, at each time slice $t$, the influence of each one of these 378 pairs of explanatory factors.

Among these $406 = (378 + 28)$ groups of explanatory factors, at each time $t$, we retain only those having both conditional probabilities of correct predictions $ (p^1 , p^0)$ larger than $  r\% $. Here $r\% > 70 \% $ is the user selected ``floor predictive power''. Note that each  single factor or pair of factors retained  at time $t$ can predict current degraded performance degradations  with a false alarm rate $FAR = 1 - p^0$ inferior to $ (100 - r) \%$.

Among the retained groups of explanatory factors, we compute the maximal influence $\max J_t$ achievable at time $t$. We also determine the {set $ \mathcal{D}_t $ of dominating groups of explanatory factors}, defined as the groups of 1 or 2 factors having an influence equal to $\MaxJ_t$ and achieving the minimal false alarm rate $(100 -r)\%$.


\subsection{Predictive power of market descriptors}

\begin{figure}[h!]
\centering
\includegraphics[width=\linewidth]{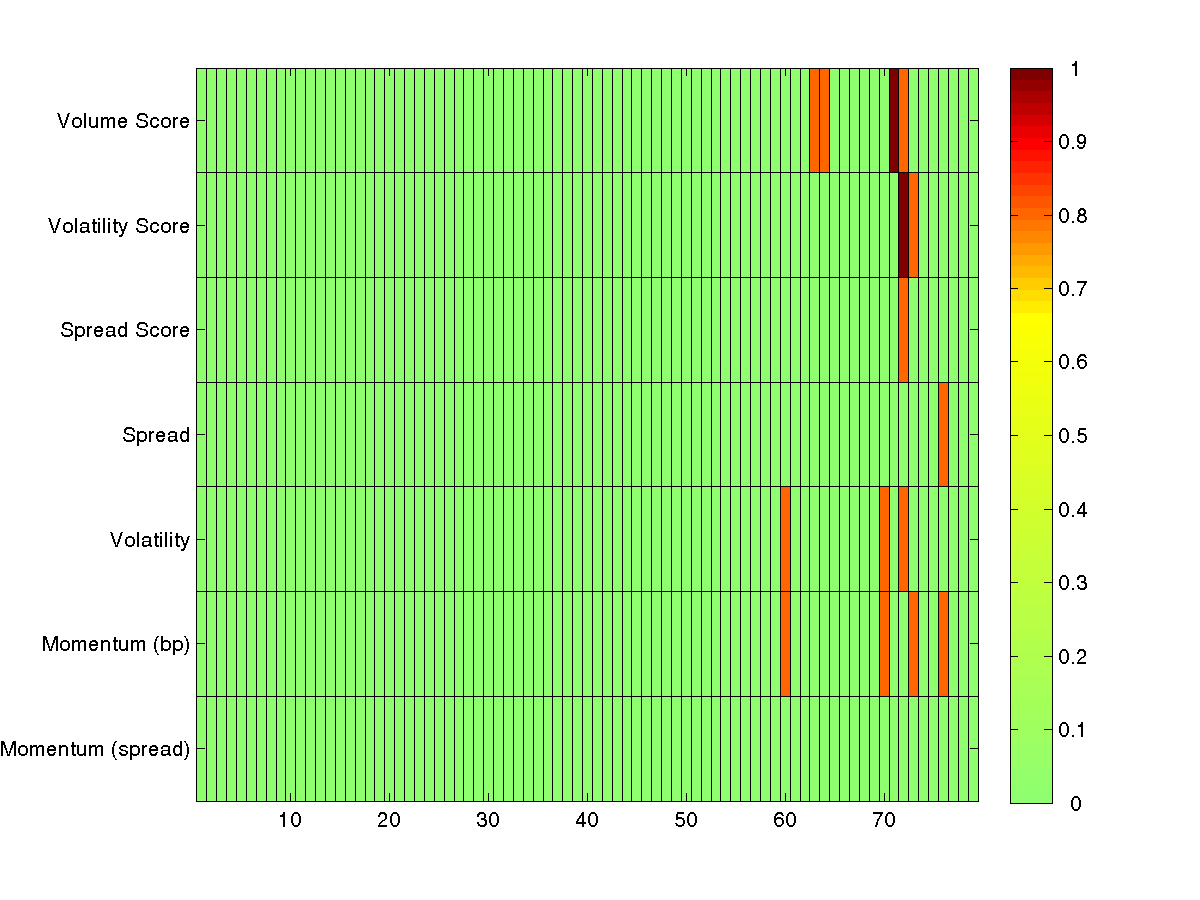}
\caption{Predictive powers of some explanatory variables (horizontal scale is time in slices of 5 minutes).}
\label{fig:explanHM}
\end{figure}

Figure \ref{fig:explanHM} gives an heatmap of the predictive power of a selected subset of market descriptors on the whole portfolio. The display shows that:
\begin{itemize}
\item no market descriptor is used before slice 60 (i.e. 14:00), meaning that there are no significant predictive links between bad trading performances and specific values of the descriptors.
\item Then the \emph{Volume Score} has the capability to explain bad trading performance from 14:10 to 14:20 and from 14:50 to 15:00.
  It means that during these two time intervals, bad performances occurred simultaneously with quite unusual levels of  traded volumes.
\item The \emph{Volatility Score} emerges as  a complementary explanatory factor between 14:55 and 15:05; orders with bad trading performances focussed on stocks having unexpectedly high volatility levels during these 10 minutes.
\item The \emph{Bid-Ask Spread Score} conforts this automated diagnosis: a rare event did indeed degrade trading performances around 15:00. 
  Keeping in mind that scores are computed according to historical values during last weeks, it means that for this portfolio, the worst performances occurred on stocks for which volumes, volatility, and bid-ask spread had abnormal values.
\end{itemize}

It is interesting to note that, before scoring, market performances  do not explain that well bad performances.


\subsection{How alarm zones explain bad trading performance}

Since the two-sided binary predictors are built to explain the degraded performances of the worst trading orders, it is easy to identify the most impacted orders one given market descriptor. Here we consider the trading order $\TO(139)$ whose lifecycle is shown on Figure \ref{fig:perfheatmapscore}, in order to visualize the impact of its \emph{Volume Score}, \emph{Volatility Score}, and \emph{Bid-Ask Score} on the order performance.
First note that this order has been active from 10:55 to 15:30 London time, with a start time two hours after the launch of the portfolio. It means that 5 minutes slices on this order are numbered from 0 to 49; they have to be shifted by 30 to be synchronized with the time scale of the other portfolio orders.

\begin{figure}[ht]
\centering
\includegraphics[width=.8\linewidth]{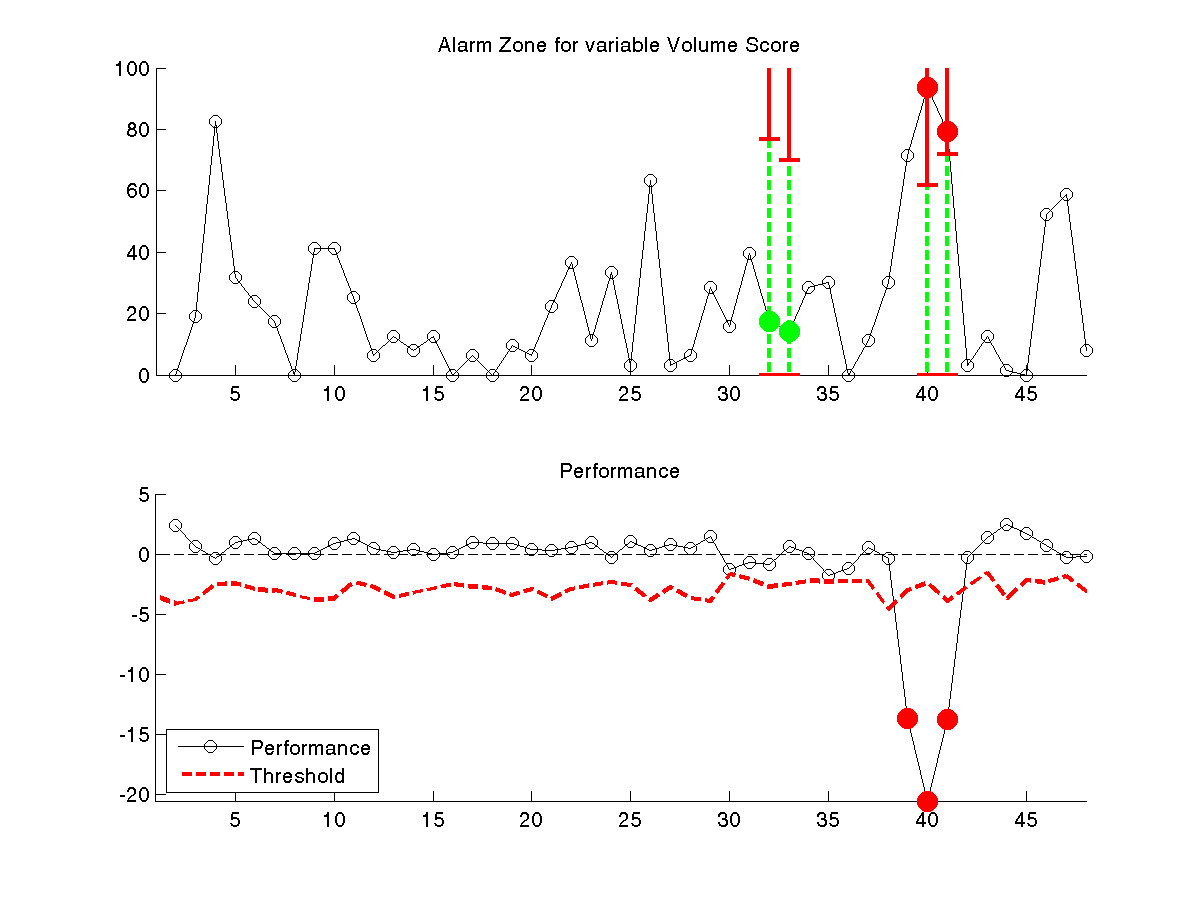}
\caption{Auto adaptive alarm zones on the \emph{Volume Score } explanatory variable for the order $T(139)$ displayed in Figure \ref{fig:perfheatmapscore}; Top: four alarm zones are active, two realizations of the Volume Score exceed the auto adaptive thresholds and thus emerge as a highly likely explanation for the  bad performance exhibited by this order (see Bottom graph.).}
\label{fig:volumezones}
\end{figure}

\paragraph{Alarm zones on the Volume Score.}
Figure \ref{fig:volumezones} shows alarm zones for the \emph{Volume Score} of order $\TO(139)$:
the top subplot draws the value of the volume score through time and the associated alarm zones have been added on top of it, when triggered.

For the whole portfolio (Figure \ref{fig:explanHM}), alarm zones are triggered on the \emph{Volume Score} from 14:10 to 14:20 and from 14:50 to 15:00 (i.e. slices 32 to 32 and 40 to 41 for this specific order). They are drawn like ``gates'' from the low threshold ($\theta^-_t$) to the high one ($\theta^+_t$); if the value of the Volume Score is outside bounds for the given order: we thus state that ``\emph{the Volume Score contributed to the degradation of trading performance for order $\TO(139)$}''.

It is important to note that if if our thresholds had not been adaptive as proposed by our ``\emph{influence analysis}'' methodology, they would have generated false explanations at the start of the order (around slice 4, i.e. 11:15 London time).

Specifically for this order, the Volume Score does not enter the first alarm zone (around 14.15); it is in line with the performance of order $\TO(139)$ that is normal (bottom subplot of Figure \ref{fig:volumezones}).
The boundaries of the next alarm zone (around 15:00) are crossed by the order Volume Score; indeed this order performs quite poorly at that time.


%
\begin{figure}[ht]
\centering
\includegraphics[width=.8\linewidth]{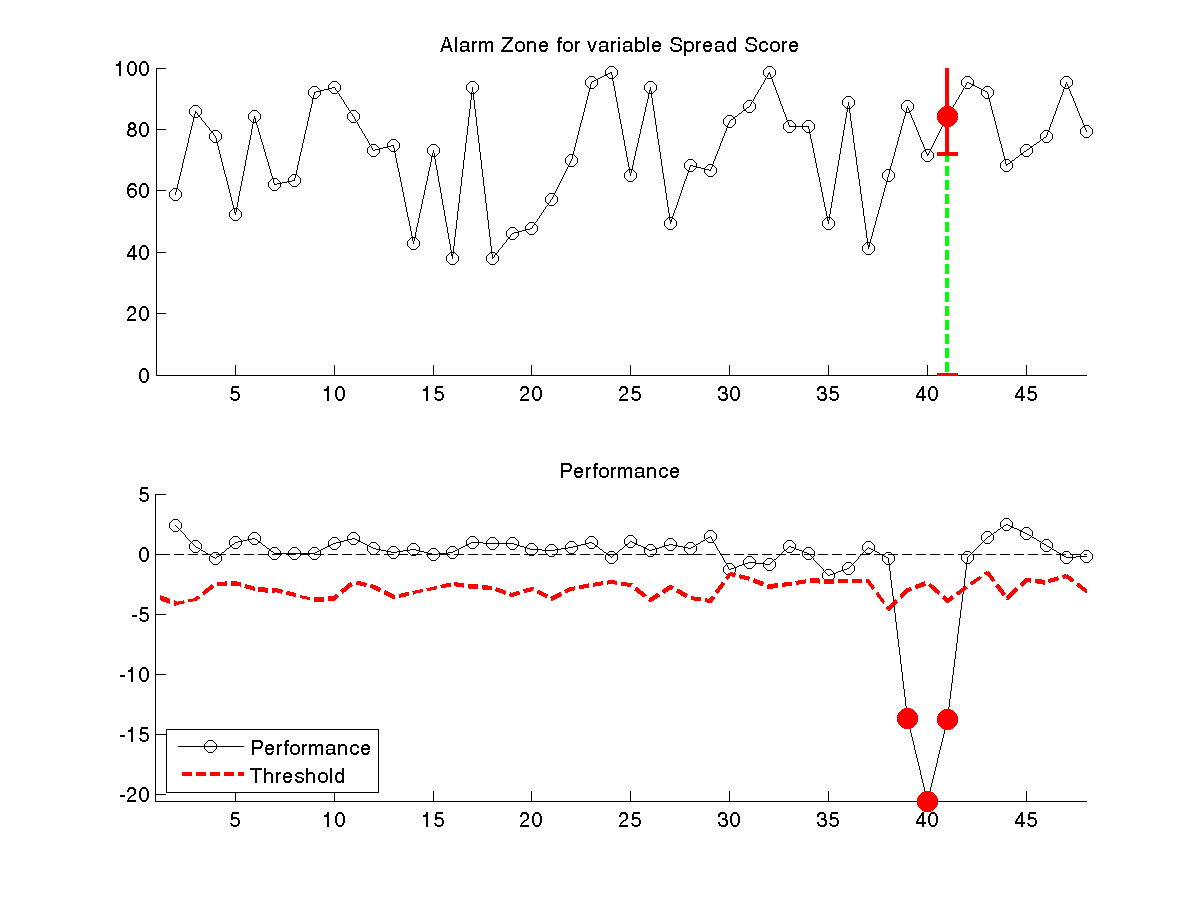}
\caption{Auto adaptive alarm zones on the \emph{Bid-Ask spread Score } explanatory variable for the order of Figure \ref{fig:perfheatmapscore}; Top: one alarm zone is active, since the Bid-Ask Score exceeds the auto adaptive thresholds, and thus emerges as a quite likely explanation of bad trading performance ( as validated  on the bottom figure).}
\label{fig:spreadzones}
\end{figure}

\paragraph{Alarm zones on the Bid-Ask Spread Score.}
Only one alarm zone has been activated on the portfolio (around 15:00) and the advantage of the auto adaptive approach proposed in this paper is straightforward: a unique threshold for all the duration of the portfolio would clearly not have been able to separate the 41st slice of order $\TO(139)$ from the others.

\begin{figure}[ht]
\centering
\includegraphics[width=.8\linewidth]{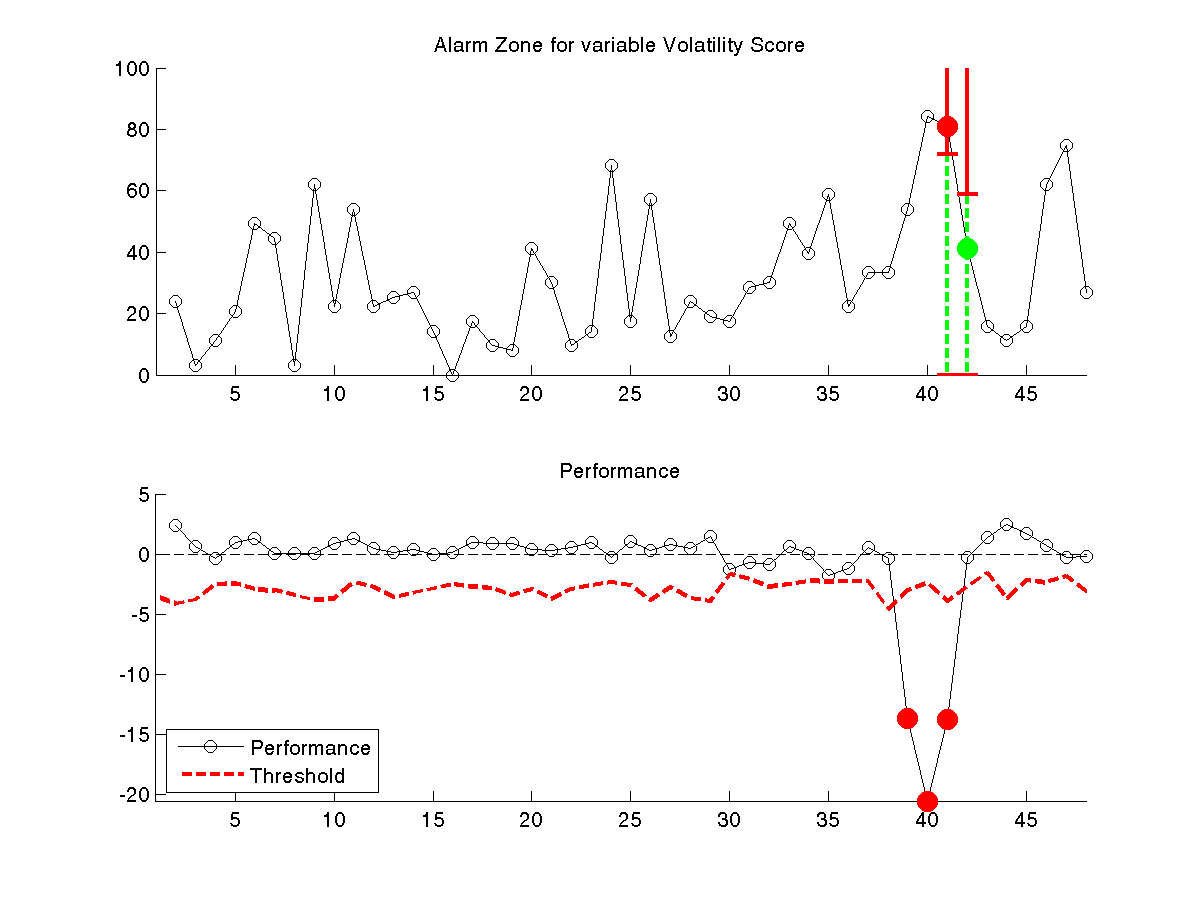}
\caption{Auto adaptive alarm zones for the \emph{Volatility Score } explanatory variable for the order of Figure \ref{fig:perfheatmapscore}; Top: two alarm zones are active, one realization of the Volatility Score exceeds the auto adaptive thresholds giving and thus provides a highly likely explanation of the current bad trading performance (as displayed on the bottom graph).}
\label{fig:volzones}
\end{figure}

\paragraph{Alarm zones on the Volatility Score.}
Once again it is clear that the alarm zones succeeded in isolating slices to efficiently explain the bad trading performance around 15:00.
\bigskip

\paragraph{To summarize.}
When applied  applied to our real portfolio of 1037 orders traded during 6 hours and 35 minutes, the \emph{automated influence analysis methodology} presented and studied from a theoretical viewpoint in this paper efficiently selects quite  pertinent explanatory  factors for degraded trading performance:
\begin{itemize}
\item our alarm zones use thresholds that are automatically adapted online to successive time slices, as computed via the \emph{predicting power} of \emph{two-sided binary predictors} (see Definition \ref{def:2pred}) based on \emph{market descriptors}.
\item Our approach generates generates auto adaptive thresholds taking into account currently observed synchronicity  between the user selected performance criterion (chosen according to the trading goal, see Section \ref{sec:tprocess}) with market descriptors \emph{market descriptors}.
\item At each time slice, the computed adaptive thresholds on market descriptors apply to the whole portfolio, bad trading performance of orders for which market descriptors take values outside \emph{alarm zones} are said to be explained or \emph{influenced} by the given descriptors. We commented real examples to illustrate these automated selection of explanatory factors.
\item The added value of augmenting the state space of market descriptors using \emph{scores} has been illustrated on several examples within our benchmark data set o trading orders..
\end{itemize}


%
\section{Conclusions}
The approach presented in this paper quantifies in real-time, the negative \emph{influence} (on trading performances) of currently detected abnormal behavior of market factors. 
The paper presents a theoretical framework to be used for TCA (Transaction Cost Analysis) and uses it for online TCA.
Dynamic {influence}-based ranking of market factors provides in real-time the most likely causality links between current bad trading performance and currently detected abnormal behavior of market variables.
To accurately capture the effect of anomalies detected on the dynamics of market variables, binary performance predictors based on anomalies have been extended to cover lagged recent occurrences of anomalies. In particular, this captures and quantifies the predictive power of crenels, jumps over multiple time steps, etc.
Our algorithms provide real-time evaluations for the current influence of specific market variables on currently observed trading performance degradations, with a small time-delay dependent on market liquidity. Moreover, \emph{influence analysis} can generate efficient real-time answers to online queries by multiple traders as well as post-trade analysis.
Our methodology enables fast in-depth dynamic evaluation of trade scheduling algorithms, and should help to quantify the comparative analysis of various trading algorithms. This will enhance new automated approaches to optimize the parameters of trading algorithms by intensive testing on historical data.
Note that we do not address here the issue of picking potential explanatory market factors, possibly from a very large pool of available factors. We consider here 7 basic market variables as given arbitrarily, or as pre-identified based on expert knowledge. However, new variables could be automatically included as potential explanatory factors by scanning a large set of market variables and ranking them after computation of their respective influence on performance degradation. For instance, it is advantageous to include factors which are uncorrelated and have historically shown to provide reliable explanation for performance degradation. 
%
%
\clearpage
\bibliographystyle{apalike}
\bibliography{lehalle-influence-paper}
\end{document}